\documentclass[12pt]{article}
\pdfoutput=1
\usepackage[T1]{fontenc}
\usepackage{baskervald}
\usepackage[font=small, labelfont=bf]{caption}
\usepackage{graphicx}
\usepackage{hyperref}
\usepackage{amsmath}
\usepackage{amssymb}
\usepackage{amsthm}
\usepackage{textcomp,bbm}
\usepackage{longtable}
\usepackage{verbatim}
\usepackage{cite}
\usepackage{enumerate}

\newcommand{\tmop}[1]{\ensuremath{\operatorname{#1}}}

\usepackage[usenames,dvipsnames]{xcolor}

\theoremstyle{definition}

\newcommand{\be}{\begin{equation}}
\newcommand{\ee}{\end{equation}}
\newcommand{\ba}{\begin{align}}
\newcommand{\ea}{\end{align}}
\newcommand{\beq}{\begin{eqnarray}}
\newcommand{\eeq}{\end{eqnarray}}

\def\simlt{\stackrel{<}{{}_\sim}}
\def\simgt{\stackrel{>}{{}_\sim}}
\usepackage{titlesec}

\newcommand{\Gam}[1]{\ensuremath{\Gamma\!\left({#1}\right)}}

\title{Sharpening the Weak Gravity Conjecture with Dimensional Reduction}

\author{Ben Heidenreich, Matthew Reece, and Tom Rudelius\\
{\small \color{gray} \texttt{bjheiden, mreece, rudelius~(@physics.harvard.edu)}}\\
Department of Physics, Harvard University, Cambridge, MA, 02138}

\begin{document}
\maketitle

\begin{abstract}
We investigate the behavior of the Weak Gravity Conjecture (WGC) under toroidal compactification and RG flows, finding evidence that WGC bounds for single photons become weaker in the infrared. By contrast, we find that a photon satisfying the WGC will not necessarily satisfy it after toroidal compactification when black holes charged under the Kaluza-Klein photons are considered. Doing so either requires an infinite number of states of different charges to satisfy the WGC in the original theory or a restriction on allowed compactification radii. These subtleties suggest that if the Weak Gravity Conjecture is true, we must seek a stronger form of the conjecture that is robust under compactification. We propose a ``Lattice Weak Gravity Conjecture'' that meets this requirement: a superextremal particle should exist for every charge in the charge lattice. The perturbative heterotic string satisfies this conjecture. We also use compactification to explore the extent to which the WGC applies to axions. We argue that gravitational instanton solutions in theories of axions coupled to dilaton-like fields are analogous to extremal black holes, motivating a WGC for axions. This is further supported by a match between the instanton action and that of wrapped black branes in a higher-dimensional UV completion.
\end{abstract}

\section{Introduction}

Only a small fraction of consistent low-energy effective quantum field theories are thought to have the potential to be consistently coupled to quantum gravity. In fact, apart from some special cases with a supersymmetric moduli space, quantum gravity theories are typically viewed as isolated points in theory space. But from the low-energy effective theory viewpoint, {\em any} quantum field theory with a conserved stress tensor can be straightforwardly coupled to gravity. The constraints imposed by quantum gravity, then, are consistency conditions that go beyond effective field theory. Low-energy field theories that cannot be completed into gravitational theories are said to reside in the ``Swampland,'' and a few heuristic criteria for determining that a theory is in the Swampland have been proposed \cite{Vafa:2005ui,ArkaniHamed:2006dz,Ooguri:2006in}. In this paper we are concerned with the Weak Gravity Conjecture (WGC) \cite{ArkaniHamed:2006dz}, which is perhaps the most useful of the Swampland criteria considered in the literature so far. For example, it places certain theories of large-field inflation in the Swampland \cite{Rudelius:2014wla,Bachlechner:2014gfa,delaFuente:2014aca,Rudelius:2015xta,Montero:2015ofa,Brown:2015iha,Bachlechner:2015qja,Hebecker:2015rya,Brown:2015lia,Junghans:2015hba,Heidenreich:2015wga,Palti:2015xra}.

Perhaps the most well-understood criterion for placing a theory in the Swampland is that a theory of quantum gravity should have no global symmetries \cite{Banks:1988yz,Kallosh:1995hi,Susskind:1995da,Banks:2010zn}. The WGC sharpens this qualitative statement into a quantitative one: an extremely weakly coupled gauge theory looks approximately like a global symmetry, and so should be more constrained than a more strongly coupled gauge theory. In four-dimensional Einstein--Maxwell theory, the WGC states that a charged particle of mass $m$ and charge $q$ should exist satisfying $m \leq \sqrt{2} q e M_{\rm Pl}$. There is also a dual statement, that a magnetic monopole of mass $m_{\rm M}$ and magnetic charge $q_{\rm M}$ should exist satisfying $m_{\rm M} \leq \sqrt{2} \frac{q_{\rm M}}{e} M_{\rm Pl}$. Identifying the classical monopole radius $r_{\rm cl} \propto (e^2 m_{\rm M})^{-1}$ with a cutoff on the validity of local effective field theory, this implies that attempting to take $e \to 0$ also sends the cutoff of the theory to zero energy, forbidding global symmetries.

We take the sharp statement of the WGC to be that, for any subextremal charged black hole in the theory, there must exist a charged particle (or a collection of charged particles) that it is kinematically possible for the black hole to emit (perhaps only marginally). For 4D Reissner-Nordstr\"om black holes this gives the bound stated in the previous paragraph. But for black holes in theories with different numbers of spatial dimensions, or in theories with massless dilaton fields that couple to the gauge field, the extremal black hole solutions will be different and the numerical coefficient in the WGC can change. In this paper we will study a variety of extremal black hole solutions in order to precisely state the WGC, including its coefficient. We then use toroidal compactification to explore the consistency of this statement. While the bound is unchanged on dimensional reduction of any single $U(1)$, we find that mixing with the resulting Kaluza-Klein photons complicates the picture, motivating a stronger form of the WGC. We also argue that certain gravitational instanton solutions in theories of axions obey an extremality bound that is closely analogous to that for black holes, motivating a version of the Weak Gravity Conjecture for axions.

Our study of compactification is similar in spirit to the earlier work \cite{Brown:2015iha}, which used $T$-duality and the M-theory limit of Type IIA string theory to explore the connection between weak gravity statements in different numbers of dimensions and for forms of different rank, notably advocating a 0-form version of the Weak Gravity Conjecture. In this paper we study toroidal compactifications of a generic Einstein-Maxwell-dilaton theory without direct reference to particular string theory realizations. As string theory is expected to always satisfy the Weak Gravity Conjecture and any consistent strengthening of it, this approach is helpful for examining how the WGC could be violated after compactification, which is central to our arguments.

\subsection{Overview of results}

Because the subsequent derivations will involve a number of technical details of black hole solutions, it is useful to first collect all of our results and discuss the physical consequences.
 In \S \ref{sec:wgc} we pin down the precise form of the Weak Gravity Conjecture for general $p$-form gauge fields in $d$ dimensions with varying dilaton couplings.  In general,
 given an abelian $p$-form with gauge coupling $e$, the WGC demands the existence of a charged $(p-1)$-brane of tension $T$ and integer charge $Q$ such that
\be
8 \pi G \left[\frac{\alpha^2}{2} + \frac{p(d-p-2)}{d-2}\right] T^2 \leq e^2 Q^2. \label{eq:introducingwgc}
\ee
Here $G$ is the $d$-dimensional Newton's constant and $\alpha$ is the coupling of a massless dilaton to the gauge theory field strength, of the form $\mathcal{L}_{\rm kin} \sim e^{-\alpha \phi} F^2$ (with conventions we will make precise below). Notice that the prefactor in this equation is $d$-dependent. When compactifying a gauge theory, the volume modulus of the compactification becomes a light mode that alters the black hole solutions, effectively changing $\alpha$ in the above formula. We show that the change in $\alpha$ compensates the change in the second term, so that the WGC is preserved under toroidal compactification, accounting for radion modes in the limit of an exactly flat moduli space. If the radions are stabilized, the WGC becomes monotonically weaker in the infrared, i.e.~the charged objects that satisfied the WGC in the higher-dimensional theory will continue to do so in the lower-dimensional theory. 

This first result is encouraging. It suggests that, if one can check that the WGC is satisfied in some theory, it will be satisfied in a range of infrared deformations of that theory. This is consistent with the expectation that the WGC should constrain infrared physics even in the absence of UV data.
With this established, we proceed to consider two subtle issues that are largely independent of each other. In \S\ref{sec:KKphotons} we examine the weak gravity consequences of mixing with the KK photon which inevitably appears in a circle compactification, whereas in \S\ref{sec:axions} we explore the extent to which the WGC might apply to axion-like fields. These sections may be read in either order. \S\ref{sec:LWGC} formulates and offers evidence from string theory for a stronger version of the WGC, motivated by our results from \S\ref{sec:KKphotons}.

We begin \S\ref{sec:KKphotons} by considering a pure gravity theory compactified on a circle. The lower dimensional theory contains a Kaluza-Klein $U(1)$ gauge field arising from graviton modes with one leg on the circle: $B_\mu \sim g_{\mu d}$. The gauge coupling becomes weaker as the circle grows larger, $e_{KK}^2 R^2 = 16 \pi G$. The charged particles in this gauge theory are the Kaluza-Klein modes of the radion and graviton, with integer charge $q$ and mass $|q|/R$. We show that, taking into account the radion coupling to the $U(1)$ field strength, these Kaluza-Klein modes all precisely saturate the Weak Gravity Conjecture bound (in the limit that the radion is not stabilized). In other words, they have the same charge-to-mass ratio as extremal black holes. Again, this result is encouraging at first glance: the Weak Gravity Conjecture has passed another necessary test, since Kaluza-Klein gauge theories automatically (though marginally) obey the bound. In fact, a second test is passed as well: the Kaluza-Klein monopole marginally satisfies the magnetic version of the Weak Gravity Conjecture.

Up to this point we have considered only a single $U(1)$ at a time. The Weak Gravity Conjecture is known to become stronger in the presence of multiple $U(1)$ gauge groups \cite{Cheung:2014vva,Heidenreich:2013}. When we compactify a $U(1)$ gauge theory on a circle, as we did in \S\ref{sec:wgc}, we also obtain the Kaluza-Klein $U(1)$ as well as an axion arising from the Wilson loop around the circle. In \S\ref{subsec:blackholebothcharge}, we construct black hole solutions labeled by the charge $Q_F$ under the original $U(1)$ and the charge $Q_H$ under the KK $U(1)$. To obtain the solutions, we lift an ordinary dilatonic charged black hole solution to one higher dimension to obtain a black string, then boost in the extra dimension, then compactify back down. The black hole solutions that we find obey an extremality bound of the form
\be
M_{\rm BH}^2 \geq \gamma e^2 M_d^{d-2} Q_F^2 + \frac{1}{R^2} \left(Q_H - \frac{\theta}{2\pi} Q_F\right)^2, \label{eq:introducing2chargeextremality}
\ee
where $\gamma$ is a constant related to the expression in brackets in (\ref{eq:introducingwgc}) and $\theta$ is the asymptotic value of the axion field. We see that the axion leads to mixing between the two $U(1)$ gauge groups. This is expected on simple physical grounds: the Kaluza-Klein charge comes from momentum along the circle direction, but for charged fields we should consider the gauge-invariant momentum from the covariant derivative $D_M = \partial_M - i Q_F A_M$. Taking the index $M$ along the circle leads to the linear combination $Q_H - \frac{\theta}{2\pi} Q_F$ that appears in the extremality bound.

Because the charges in (\ref{eq:introducing2chargeextremality}) add in quadrature---as is familiar from simpler multi-charge black hole solutions---the existence of particles satisfying WGC for the fundamental $U(1)$ and for the KK $U(1)$ {\em separately} does not guarantee that an extremal black hole carrying {\em both} charges will be able to decay. The criterion that guarantees this is the convex hull condition (CHC) \cite{Cheung:2014vva}. Labeling the charged particles with an index $i$ and the collection of $U(1)$ gauge groups with the index $a$, we consider the charge-to-mass vectors
\be
{\vec \zeta}_i = \frac{1}{m_i} \left(q_{i1}, q_{ia}, \ldots q_{iN}\right).
\ee
In a theory without dilaton-like fields, the CHC says that the convex hull of the collection of vectors $\{\pm {\vec \zeta}_i\}$, with $i$ ranging over all charged particles in the theory, should contain the unit ball as measured with a metric determined by the kinetic matrix of the gauge fields. This metric is the same quadratic form that appears in the appropriate black hole extremality bound; in the simplest case of a collection of unmixed $U(1)$s, this quadratic form is
\be
\left<{\vec \zeta}_i, {\vec \zeta}_j\right> = M_d^{d-2}  \sum_a \gamma_a e_a^2 \zeta_{ia} \zeta_{ja}.
\ee 
From this we immediately see that for a collection of $N$ gauge groups with equal gauge couplings and spectra of charged particles the WGC bound becomes stronger by a factor of $\sqrt{N}$ \cite{Cheung:2014vva}, a fact that has important consequences for the idea of $N$-flation \cite{Rudelius:2014wla,Heidenreich:2015wga}.

Once we take scalar fields with dilaton-like couplings into account, the CHC is replaced by a slightly different statement: rather than the unit ball computed with some metric, the statement is simply that the convex hull of the vectors ${\vec \zeta}$ must contain the region of all ${\vec q}/m$ values attained by subextremal semiclassical black holes. We show that in some cases this region has a polygonal boundary, rather than an ellipsoidal one, so the presence of scalar fields marks a significant difference in the form of the convex hull condition. Concrete examples in string theory require this modified statement of the CHC.
 
For now we are interested in the implication of the CHC for the scenario of two $U(1)$s, one of which we can view as fundamental and one of which arises from Kaluza-Klein reduction on a circle. The danger is apparent: because the KK $U(1)$'s own WGC bound is only {\em marginally} satisfied by the KK graviton, we run the risk that a black hole charged under a combination of the two $U(1)$s will not be able to emit any charged particles. In fact, this is our conclusion: if the WGC is satisfied for a $U(1)$ gauge theory in $d$ dimensions, its CHC cousin for the dimensionally reduced theory in $d-1$ dimensions is {\em not} necessarily satisfied. This is the first roadblock that we have hit in attempting to check the internal consistency of the WGC.

To understand what this surprise is telling us, let us consider some ways that the lower-dimensional CHC could be satisfied. As one example, consider a theory compactified on a torus. In this case, the KK modes on each circle independently saturate their respective WGC bounds. The case where each WGC is only marginally satisfied seems particularly dangerous for the CHC. But a Kaluza-Klein compactification on a torus actually {\em does} satisfy the convex hull condition. There is a set of graviton modes with arbitrary integer charges $(n_1, n_2)$ under the two $U(1)$s. They have mass $\sqrt{(n_1/R_1)^2 + (n_2/R_2)^2}$ and as such marginally satisfy the relevant extremality condition for any direction in the charge lattice. In this particular case, a {\em much} stronger form of the WGC than we usually consider is true: there is an independent single-particle state satisfying WGC for every direction in the charge lattice. We speculate that  this very strong form of the WGC may actually be true in all string theory constructions. It has been conjectured in the past that quantum gravity demands that there is a state in the theory corresponding to any point in the charge lattice allowed by Dirac quantization \cite{Polchinski:2003bq,Banks:2010zn}; perhaps an even stronger statement is true, that there is an extremal or superextremal state for every point in the charge lattice. We call this possibility the Lattice Weak Gravity Conjecture. This strengthens a conjecture proposed in \cite{Brown:2015iha}, which requires that the lightest state in any direction in the charge lattice be superextremal. At first glance it appears to contradict a statement made in \cite{ArkaniHamed:2006dz} about the spinor of $SO(32)$ in heterotic string theory; we argue in \S\ref{sec:SO32spinor} that it does not---in fact, we show that the weakly coupled heterotic string satisfies the Lattice Weak Gravity Conjecture! As a further suggestive piece of evidence, it has been shown in some cases that corrections from higher-dimension operators suppress the mass-to-charge ratio of extremal black holes \cite{Kats:2006xp}. Thus, if there is an extremal or superextremal state for every charge in the charge lattice, for very large charges these might be approximately semiclassical black hole states, while for very small charges they could be ordinary particles or low-lying string modes.

There are other ways that the CHC could be satisfied. The first is that the WGC in the higher-dimensional theory could be satisfied by a single charged particle that obeys the bound with some room to spare. The infinite tower of KK modes of this single particle have a combination of the two $U(1)$ charges, and it is possible for their convex hull to contain the unit ball. We illustrate this possibility in Figure \ref{convexhull}. We note that this possibility is unavailable if we demand that the CHC be satisfied in the compactified theory for {\em any} radius $R$. It is available only if there is a minimal available radius $R_{\rm min}$, possibly corresponding to the distance at which effective field theory breaks down. This possibility suggests that we will need to impose a stronger condition than the WGC in the higher-dimensional theory, but perhaps a different one than an infinite set of constraints on every direction in the charge lattice. Alternatively, we note that the KK $U(1)$ marginally satisfies the WGC only when we take the radion to be massless. In many theories the radion will be stabilized, weakening the relevant WGC bound. This provides another possible way that the CHC could be satisfied, though if we consider black holes of size smaller than the radion Compton wavelength it is not clear that this route is available. In any case, it is clear that consideration of the CHC in the Kaluza-Klein context reveals previously overlooked subtleties in the interpretation of the Weak Gravity Conjecture. We expect that these subtleties point in fruitful directions for further research.

In \S\ref{sec:axions} we examine the extent to which the WGC can be extended to axions. The original WGC paper suggested an analogous axion conjecture, viewing the axion as a 0-form gauge field, instantons as the analogue of charged objects, and the instanton action as the analogue of mass or tension. From the coefficient in (\ref{eq:introducingwgc}), an equation that was derived only for $1 \leq p \leq d-3$, we see that a na\"ive extrapolation to $p = 0$ breaks down unless $\alpha \neq 0$. We will present two lines evidence that lead us to believe that the $p = 0, \alpha \neq 0$ bound may actually apply to axions in theories of quantum gravity. The first is intrinsic to the axion theory. While there are no analogues of extremal charged black holes in a theory of axions, there is a class of Euclidean gravitational instanton solutions (recently receiving attention in the WGC context~\cite{Montero:2015ofa,Bachlechner:2015qja,Brown:2015lia}). In fact, both wormhole and instanton solutions have been discussed in the literature. We argue that it is only the instantons that are of interest to us: they contribute to the axion potential and have a clear interpretation in ordinary local field theory, unlike wormholes whose status in quantum gravity is murkier. One source of confusion in the literature is that if there is no dilaton one finds only wormhole solutions, rather than instantons. But this is perfectly consistent with the extrapolation of (\ref{eq:introducingwgc}), which leads us to expect that the bound degenerates when a dilaton is not present. One way to think about the instanton solutions is that a curvature singularity develops at a finite distance down the throat of a would-be wormhole, effectively pinching it off and producing a purely local effect. Despite the singular nature of the solutions, we show that for $\alpha$ larger than a critical value there is a well-defined instanton action obeying an extremality bound that is precisely analogous to the $p \to 0$ limit of the extremal black hole bound.

Our second line of evidence comes from considering axions that arise from dimensionally reducing higher-rank $p$-form gauge fields. This is a standard way to obtain axion-like fields with controlled violations of shift symmetry. In this case, the ordinary WGC applies to the original gauge field, so there are charged objects in the higher-dimensional theory. Wrapping the worldvolume of these charged objects around the cycle on which we dimensionally reduce leads to an instanton effect in the axion effective potential. We show that if the wrapped object is a black hole, the instanton action in the dimensionally reduced theory precisely corresponds to that of the gravitational instanton solution we derived purely in the lower-dimensional theory. The range of $\alpha$ over which the instanton extremality bound has the expected form is precisely the range that is obtainable via dimensional reduction.

It has recently been argued that gravitational instantons can satisfy the axion WGC while providing negligible contributions to the axion potential, taking advantage of possible loopholes in WGC arguments pointed out in \cite{Rudelius:2015xta,Montero:2015ofa,Brown:2015iha} in a manner that would effectively decouple WGC constraints from inflationary model-building \cite{Bachlechner:2015qja}. In light of our results, we believe that this is not the correct interpretation of the gravitational instantons. Gravitational instantons play a role precisely analogous to the role that black holes play in the case of the ordinary WGC, and the ordinary WGC is generally believed to be stronger than the simple statement that extremal black holes exist. To the extent that the WGC is generally taken to imply that charged objects that are {\em not} extremal semiclassical black holes exist in a theory, the analogous WGC for axions should imply that instanton effects that are {\em not} simply extremal gravitational instantons should exist. These are expected to have smaller instanton action and hence to play a dominant role in determining the axion potential. Similarly, if the Lattice WGC is true, then the extremal gravitational instantons play the role of extremal black holes, filling out the large-charge regions of the charge lattice. But the small-charge regions must still be filled by instantons that are not well approximated by gravitational instantons, just as the small-charge points in the charge lattice of a $U(1)$ gauge theory are occupied by low-lying particle or string states.

Without further ado, let us turn to the detailed version of the arguments and calculations we have just summarized. We will offer a few concluding remarks and thoughts on the next steps to pursue in \S\ref{sec:outlook}.

\section{Weak Gravity for p-branes and circle compactifications}
\label{sec:wgc}

\subsection{Conventions}

Before we address physics, let us fix our conventions. We will work in a mostly-plus signature for the metric. We take the action for general relativity coupled to a dilaton field and a $p$-form gauge field in $d$ dimensions to be
\be
S = \frac{1}{2\kappa_d^2} \int d^d x \sqrt{-g} \left({\cal R}_d - \frac{1}{2} (\nabla \phi)^2\right) - \frac{1}{2e_{p;d}^2} \int d^d x \sqrt{-g} e^{-\alpha_{p;d} \phi} F_{p+1}^2 \,. \label{eq:generalaction}
\ee
Note that the dilaton in this expression is
not canonically normalized; we will refer to it as 
``conventionally normalized.'' Here $F_{p+1} = d A_p$ is the field strength for a $p$-form gauge field $A_{\mu_1 \ldots \mu_p}$, with
\be
F_q^2 \equiv \frac{1}{q!} F_{\mu_1 \ldots \mu_q} F^{\mu_1 \ldots \mu_q} \,.
\ee
With this definition, $\star F_q^2 = F \wedge \star F$. The gauge field $A_{\mu_1 \ldots \mu_p}$ has dimension $p$, so that the integral $\int_{\Sigma_p} A_p$ over the worldvolume of a charged $(p-1)$-dimensional object is dimensionless. The coupling constant $e_{p;d}^2$ has dimension $2(p+1)-d$ whereas $\alpha_{p;d}$ is dimensionless. The Ricci scalar ${\cal R}_d$ has dimension 2, so $\kappa_d^2$ has dimension $2-d$. We will also use the reduced $d$-dimensional Planck mass $M_d$ and Newton constant $G_d$ defined by
\be
\frac{1}{\kappa_d^2} = \frac{1}{8 \pi G_d} = M_d^{d-2}.
\ee
The subscripts $p$ and $d$ are useful when matching theories in different dimensions, but will sometimes be dropped for convenience.
In the case of four dimensions we will sometimes write $M_{\rm Pl}$ rather than $M_4$.

We denote the volume of a unit $d$-dimensional sphere $S^d$ by $V_d$ and the corresponding volume form by $\omega_d$, i.e.
\be
V_d = \int_{S^d} \omega_d = \frac{2 \pi^{\frac{d+1}{2}}}{\Gam{\frac{d+1}{2}}}. \label{eq:sphereV}
\ee
The electric charge of a $(p-1)$-brane and the magnetic charge of a dual $(d-p-3)$-brane are defined by
\begin{align}
Q &= \frac{1}{e_{p;d}^2} \int_{S^{d-p-1}} e^{-\alpha_{p;d} \phi}\star F, \label{eq:Qel}\\
{\tilde Q} &= \frac{1}{2\pi} \int_{S^{p+1}} F. \label{eq:Qmag}
\end{align}
Dirac quantization implies that $Q {\tilde Q} \in {\mathbb Z}$. The coupling $e_{p;d}^2$ is chosen such that $Q$ is quantized in integer units, so that the coupling of $A_p$ to a minimally charged $(p-1)$-brane with worldvolume $\Sigma_p$ is simply $S = \int_{\Sigma_p} A_p$. For example, with these conventions the field strength of a point charge in $d$-dimensional Maxwell theory is $F = \frac{e^2 Q}{V_{d-2} r^{d-2}} dt \wedge dr$.

Under electromagnetic duality,
\be
- \frac{1}{2e_{p;d}^2} \int d^d x \sqrt{-g} e^{-\alpha_{p;d} \phi} F_{p+1}^2 \to - \frac{1}{2g_{p;d}^2} \int d^d x \sqrt{-g} e^{\alpha_{p;d} \phi} G_{d-p-1}^2,
\ee
where the magnetic coupling and field strength are given by
\beq
g_{p;d}^2 & = & \frac{4\pi^2}{e_{p;d}^2}, \nonumber \\
G_{d-p-1} & = & \frac{2 \pi}{e_{p;d}^2} e^{-\alpha_{p;d} \phi} \star F_{p+1}.
\eeq
The formulas (\ref{eq:Qel}) and (\ref{eq:Qmag}) transform into each other under this operation.

For later convenience, we define the quantity
\be
\gamma_{p;d}(\alpha) \equiv \left[\frac{\alpha^2}{2} + \frac{p (d - p - 2)}{d-2}\right]^{-1}. \label{eq:gammadef}
\ee
As we will see, this quantity sets the extremality bound for black branes, as well as playing an important role in other aspects of black brane solutions.
The identity $\gamma_{p;d}(\alpha) = \gamma_{d-p-2;d}(-\alpha)$ is a manifestation of electromagnetic duality.

\subsection{Dilatonic black branes in arbitrary dimensions}
\label{sec:dilatonicbranes}

There is an extensive literature on dilatonic black hole and black brane solutions \cite{Gibbons:1982ih,Myers:1986un,Gibbons:1987ps,Garfinkle:1990qj,Horowitz:1991cd,Lu:1993vt,Duff:1993ye,Duff:1996hp}. For our purposes, the most directly useful result is the solution of Horowitz and Strominger for dilatonic black hole solutions that have a $(d-2)$-form magnetic field strength in $d$ dimensions (equations (5) through (11) of \cite{Horowitz:1991cd}). Throughout this paper, we will repeatedly use the Horowitz--Strominger solution together with dimensional reduction or lifting to higher dimensions to generate all of the other solutions of interest to us. The technique is the same one already used to obtain a variety of solutions for black branes in 10d string theory; we will simply generalize to arbitrary $p$-forms, number of dimensions $d$, and dilaton coupling $\alpha_{p;d}$.

We seek a magnetically charged black brane solution for the action (\ref{eq:generalaction}) in $d$ dimensions. The magnetic brane worldvolume has $d - p - 2$ dimensions; we take an ansatz that is translationally and rotationally invariant in the $n=d-p-3$ spatial worldvolume coordinates $y^i$, $i = 1, \ldots n$. We can then dimensionally reduce to an $m=d-n=p+3$ dimensional theory with the metric ansatz
\be
ds^2 = e^{\frac{n}{m-2} \lambda} d{\hat s}^2 + e^{-\lambda} \delta_{ij} dy^i dy^j, \label{eq:dimredansatz}
\ee
where $d{\hat s}^2$ is an arbitrary $m$-dimensional metric and $\lambda$ is a scalar warp factor, both independent of $y^i$. The dimensionally reduced action is 
\be
\frac{1}{2\kappa^2} \int d^m x \sqrt{-{\hat g}} \left({\hat {\cal R}} - \frac{1}{2} \partial_\mu \phi \partial^\mu \phi - \frac{n(d-2)}{4(m-2)} \partial_\mu \lambda \partial^\mu \lambda\right) - \frac{1}{2 e_p^2} \int d^m x \sqrt{-\hat{g}} e^{- \alpha \phi - \frac{n (m-3)}{m-2}\lambda} F_{p+1}^2,
\ee
where indices are understood to be raised with ${\hat g}$.

At this point we use a strategy that will recur throughout this paper. In this case we write the details explicitly; in subsequent sections we will simply state the final results. The idea is to seek solutions in which the linear combination of $\phi$ and $\lambda$ that appears in the exponent coupling to $F^2$ is turned on but the orthogonal linear combination is zero \cite{Horowitz:1991cd}. Explicitly, we match to conventionally normalized fields $\rho$ (which couples to $F^2$) and $\sigma$ (which does not) via 
\begin{align}
\beta \rho  &= \alpha \phi + \frac{n (m-3)}{m-2} \lambda \,, \nonumber \\
\beta \sigma &= \frac{2(m-3) \phi - \alpha (d-2) \lambda}{d-2} \sqrt{\frac{n (d-2)}{2(m-2)}} \,, & \beta  &=  \sqrt{\alpha^2 + \frac{2n(m-3)^2}{(m-2)(d-2)}}\,. \label{eq:rediagonalize}
\end{align}
Making these replacements in equation (11) of \cite{Horowitz:1991cd} and substituting the ansatz (\ref{eq:dimredansatz}) allows us to find a solution for our original action. In terms of two functions that vanish at the outer and inner horizons respectively,
\be
f_\pm(r) \equiv 1 - \left(\frac{r_\pm}{r}\right)^p,
\ee
we have the solution
\begin{align}
ds^2 &= -f_+(r) f_-(r)^{\gamma_\parallel-1} dt^2 + f_+(r)^{-1} f_-(r)^{\gamma_\perp-1} dr^2 + r^2 f_-(r)^{\gamma_\perp} d\Omega_{p+1}^2 + f_-(r)^{\gamma_ \parallel} \delta_{ij}dy^i dy^j, \nonumber \\
e^{-\alpha \phi} &= f_-(r)^{p \gamma_\perp}, \nonumber \\
F_{p+1} &= \frac{e_{p;d}}{\kappa_d} {\cal Q} \omega_{p+1}, \label{eq:branesolution}
\end{align}
where $\gamma_\parallel \equiv \frac{2p}{d-2} \gamma$, $\gamma_\perp \equiv \frac{\alpha^2}{p} \gamma$, and ${\cal Q}^2 = p^2 (r_+ r_-)^p \gamma$ all have simple expressions in terms of the quantity $\gamma = \gamma_{p;d}(\alpha)$ defined in (\ref{eq:gammadef}). This general solution has previously been given by Duff, L\"u, and Pope~\cite{Duff:1996hp}, albeit in a different coordinate system that we find less useful.

From this result we can compute the magnetic charge and the ADM tension:\footnote{See for instance (2.8) in~\cite{Lu:1993vt} for the ADM tension of a black brane.}
\begin{align}
{\tilde Q} &= \frac{e_{p;d} V_{p+1}}{2\pi \kappa_d} p (r_+ r_-)^{p/2} \sqrt{\gamma} = \frac{V_{p+1}}{g_{p;d} \kappa_d} p (r_+ r_-)^{p/2} \sqrt{\gamma},\\
T &= \frac{V_{p+1}}{2\kappa_d^2} \left[(p+1) \left(r_+^p - r_-^p\right) + 2 p \gamma r_-^p\right].
\end{align}
The extremality bound $r_+ \geq r_-$ corresponds to the inequality
\be
\gamma g_{p;d}^2 {\tilde Q}^2 < \kappa^2 T^2. \label{eq:magneticextremality}
\ee
We can apply electromagnetic duality to obtain the analogous electrically charged black holes. The metric takes the same form as above, with $\gamma_\parallel = \frac{2(d-p-2)}{d-2} \gamma$, $\gamma_\perp = \frac{\alpha^2}{d-p-2} \gamma$, and $d\Omega_{d-p-1}^2$ in places of $d\Omega_{p+1}^2$.\footnote{As noted above, $\gamma_{p;d}(\alpha) = \gamma_{d-p-2;d}(-\alpha)$, hence there is no need to distinguish between electromagnetic duals in computing $\gamma$.} Furthermore, for electrically charged black holes,
\be
f_\pm(r) \equiv 1 - \left(\frac{r_\pm}{r}\right)^{d-p-2}\,.
\ee
The flux and dilaton profiles are now
\begin{align}
e^{\alpha \phi} &= f_-(r)^{(d-p-2) \gamma_\perp}\,, &
F_{p+1} & =  \frac{e_{p;d}}{\kappa_d} (d-p-2) \frac{\sqrt{\gamma}\, (r_+ r_-)^{\frac{d-p-2}{2}}}{r^{d-p-1}}\, \Omega_p \wedge d r \,,
\label{eq:electricsolution}
\end{align}
where $\Omega_p = dt \wedge dy^1 \wedge \ldots \wedge dy^{p-1}$ denotes the volume form along the brane.
The extremality bound becomes
\be
\gamma e_{p;d}^2 Q^2 < \kappa^2 T^2. \label{eq:electricextremality}
\ee
For instance, for $d = 4$, $p = 1$, and $\alpha = 0$, we find that $\gamma = 2$, recovering the familiar result for Reissner-Nordstr\"om black holes: $2 e^2 Q^2 M_{\rm Pl}^2 < M^2$.

\subsection{The general Weak Gravity Conjecture}

We have taken our working statement of the Weak Gravity Conjecture to be that we demand the existence of a superextremal particle or brane that allows any extremal charged black hole or black brane to decay. That is, a charged object of tension $T_p$ and quantized charge $q$ should exist that violates (\ref{eq:electricextremality}). It is useful to state this with the explicit value of $\gamma$:
\be
{\rm Weak~Gravity:}~~~\left[\frac{\alpha^2}{2} + \frac{p(d-p-2)}{d-2}\right] T_p^2 \leq e_{p;d}^2 q^2 M_d^{d-2}.
\label{eq:generalWGC}
\ee
Strictly speaking, our derivation of the extremality bound is valid only for $1 \leq p \leq d-3$, but it is interesting to examine how the formula behaves for all $p$ if we na\"ively extrapolate it. Notice that in the case $\alpha = 0$, the extremality bound becomes degenerate for $p = 0$ or $p = d-2$: it would imply that no charged black objects exist. These two cases are clearly special. Objects charged under a 0-form are instantons, which lack time evolution. Objects charged under $d-2$ forms include point particles in three dimensions, cosmic strings in four dimensions, and D7 branes in string theory: their gravitational backreaction in flat space leads to a deficit angle, and so beyond some critical tension the existence of a single such object would completely destroy the space. Thus, it is not surprising that there is no precise analogue of the black hole extremality bound for these two special choices of $p$. There are two other special cases: $p = d$, e.g.~D9 branes in string theory, is unlikely to be interesting since the net charge must be canceled; $p = d-1$, including domain walls in 4d or D8 branes in string theory, may be worth considering, but we will not discuss it here.

In the remainder of this paper, we will explore the consistency of the Weak Gravity Conjecture under dimensional reduction. Given a gravitational theory containing a $p$-form gauge field that satisfies the Weak Gravity Conjecture in $d$ dimensions, we can compactify on a circle to obtain $p$-form and $(p-1)$-form gauge fields in $d-1$ dimensions. The charged particles in this lower-dimensional theory are inherited from the higher-dimensional theory. It would be surprising if they are insufficient to satisfy WGC: that is, we (na\"ively) expect the condition that the WGC is satisfied in $d$ dimensions to be {\em stronger} than the condition that it be satisfied in $d-1$ dimensions. Indeed, when we compactify on a circle of radius $R$, both the gauge coupling and the Planck mass are related by the volume of the internal dimension:
\be \label{eqn:redPlanck}
\frac{1}{e_{d-1}^2} = \frac{2\pi R}{e_d^2},~~~M_{d-1}^{d-3} = 2\pi R M_d^{d-2}.
\ee
As a result, the bound (\ref{eq:generalWGC}), at fixed $\alpha$, becomes strictly easier to satisfy in the dimensionally reduced theory. However, the full story is slightly more subtle. Compactifications from extra dimensions will always introduce new scalar degrees of freedom. In the simplest case of a circle compactification, this is the radion mode, which plays the role of a dilaton with $\alpha \neq 0$ in the action (\ref{eq:generalaction}). When the dilaton is massless, we will see that the effective value of $\alpha$ adjusts under compactification in precisely the right way for the bound (\ref{eq:generalWGC}) to become neither weaker nor stronger. In a nonsupersymmetric setting, the radion will generally obtain a mass, so the compactified theory at long distances flows back to Maxwell--Einstein theory with no scalar mode and the Weak Gravity bound becomes the standard one again, which is a weaker bound than in the Maxwell--Einstein--radion theory. This suggests an interesting monotonicity property: Weak Gravity bounds should always become weaker as one flows toward the IR in a given theory.

The observation that compactifications lead to actions with $\alpha \neq 0$, at least over some range of distances, re-opens the question of whether it is sensible to discuss a 0-form Weak Gravity Conjecture. This case is of interest for axions in scenarios like extranatural inflation, because the 0-form we are interested in studying in four dimensions arises from a 1-form or higher in a UV completion of the 4D theory. The nontrivial bound that exists at $\alpha \neq 0$ descends from the higher-dimensional theory. In fact, even within the axion theory itself, there exists a close analogue of extremal black holes in the form of extremal gravitational instantons (when $\alpha \neq 0$). We will return to this question in \S\ref{sec:axions}.

In the next two subsections we will show that the general bound is well-behaved under compactification on a circle, either preserving or decreasing the rank of our gauge field.

\subsection{Dimensional reduction on a circle, preserving $p$} \label{subsec:dred-pconst}

Suppose that we begin with the action (\ref{eq:generalaction}) in $D$ dimensions and compactify down to $d = D-1$ dimensions on a circle of radius $R$. We parametrize the $D$ dimensional-metric $g$ in terms of the $d$-dimensional Einstein frame metric ${\hat g}$ and radion mode $\lambda$ by:
\be
ds^2 = e^{\frac{\lambda(x)}{d-2}} d{\hat s}^2(x) + e^{-\lambda(x)} dy^2.
\ee
We consider the $p$-form in $d$ dimensions descending from the $p$-form in $D$ dimensions, with the dimensionally reduced action
\be
S = \int d^d x \sqrt{-{\hat g}}\left[\frac{1}{2\kappa_d^2} \left({\cal {\hat R}}_d - \frac{1}{2} (\nabla \phi)^2 - \frac{d-1}{4(d-2)} (\nabla \lambda)^2\right) - \frac{1}{2 e_{p;d}^2} e^{-\alpha_{p;D} \phi - \frac{p}{d-2} \lambda} F_{p+1}^2\right].
\ee
All indices in this expression are raised with ${\hat g}$. One factor of $e^{-\frac{\lambda}{d-2}}$ arising from raising an index on $F$ with $g^{\mu \nu}$ is compensated by the factor in $\sqrt{-g}$, so $p$ factors remain from the other indices, giving rise to the $\lambda$ dependence in the exponent multiplying $F_{p+1}^2$.

At this point we recycle an idea from \S\ref{sec:dilatonicbranes} and \cite{Horowitz:1991cd}: we rewrite $\lambda$ and $\phi$ in terms of two conventionally normalized fields, one of which (call it $\sigma$) is decoupled from $F_{p+1}^2$ and can be set to zero in the solution while the other (call it $\rho$) couples to $F_{p+1}^2$ via $e^{-\alpha_{p;d}\, \rho}$. If we define a conventionally normalized radion via $\hat{\lambda} \equiv \sqrt{\frac{d-1}{2(d-2)}} \lambda$, its coupling to the gauge field is given by $e^{-\beta_{p;d} {\hat \lambda}} F_{p+1}^2$ where
\be \label{eqn:radioncoupling}
\beta_{p;d} \equiv  \sqrt{\frac{2 p^2}{(d-1)(d-2)}}.
\ee
A computation completely analogous to the one that led to equation (\ref{eq:rediagonalize}) tells us that 
\be \label{eqn:psame_alpha}
\alpha_{p;d}^2 = \alpha_{p;D}^2 + \beta_{p;d}^2 = \alpha_{p;D}^2 + \frac{2 p^2}{(d-1)(d-2)}
\ee
is the coupling of the effective dilaton $\rho$.
This can be rewritten as:
\be
\frac{\alpha_{p;d}^2}{2} +\frac{p (d-p-2)}{d-2} = \frac{\alpha_{p;D}^2}{2} + \frac{p (D-p-2)}{D-2} \,,
\ee
or $\gamma_{p;d}(\alpha_{p;d}) = \gamma_{p;D}(\alpha_{p;D})$.
Using (cf.~(\ref{eqn:redPlanck})):
\be
\frac{1}{e_{p;d}^2} = \frac{2\pi R}{e_{p;D}^2},~~~M_d^{d-2} = (2\pi R) M_D^{D-2} \,,
\ee
we conclude that the extremality bound~(\ref{eq:generalWGC}) is unchanged after compactification on a circle.

\subsection{Dimensional reduction on a circle, decreasing $p$} \label{subsec:decreasingp}

Now consider a slightly different case, reducing from $D$ to $d = D-1$ dimensions but also reducing the rank of the form we consider---therefore also the dimensionality of the black brane---from $P$ to $p=P-1$ via $A_{p} = \oint dy~A_P$. In this case, one of the indices on $F^2$ is raised with $g^{yy} = e^{\lambda}$ rather than $g^{ij} = e^{-\frac{\lambda}{d-2}} \delta^{ij}$, changing the exponent relative to the previous case:
 \be
S = \int d^d x \sqrt{-{\hat g}}\left[\frac{1}{2\kappa_d^2} \left({\cal {\hat R}}_d - \frac{1}{2} (\nabla \phi)^2 - \frac{d-1}{4(d-2)} (\nabla \lambda)^2 \right) - \frac{1}{2 e_{p;d}^2} e^{-\alpha_{P;D} \phi + \frac{d-p-2}{d-2} \lambda} F_{p+1}^2\right].
\ee
Again we can treat this as an effective coupling $e^{-\alpha_{p;d}\, \rho}$ of one conventionally normalized scalar field to $F_{p+1}^2$, but in this case
\be \label{eqn:preduce_alpha}
\alpha_{p;d}^2 = \alpha_{P;D}^2 + \frac{2(d-p-2)^2}{(d-1)(d-2)}.
\ee
As above, this can be rewritten as
\be
\frac{\alpha_{p;d}^2}{2} + \frac{p (d-p-2)}{d-2} = \frac{\alpha_{P;D}^2}{2} + \frac{P (D-P-2)}{D-2}\,,
\ee
or $\gamma_{p;d}(\alpha_{p;d}) = \gamma_{P;D}(\alpha_{P;D})$.
We now have 
\be
e_{p;d}^2 = (2\pi R) e_{P;D}^2 \;\;,\;\; T_p = (2 \pi R) T_P \,,
\ee
along with $M_d^{d-2} = (2\pi R) M_D^{D-2}$, so the factors of $(2 \pi R)$ cancel and the extremality bound~(\ref{eq:generalWGC}) is again unchanged.

Notice that the radion coupling which appears in~(\ref{eqn:preduce_alpha}) is the same as the coupling $\beta_{d-p-2;d}$ from~(\ref{eqn:radioncoupling})--(\ref{eqn:psame_alpha}). More generally, the arguments of this section and the previous one are related by electromagnetic duality, which exchanges form fields with and without legs along the compact circle.

Given these results, we see that the WGC is well-behaved under any toroidal compactification. Furthermore, if dilaton or radion modes acquire a mass, the bounds always become monotonically weaker in the infrared.

\section{Weak gravity and KK photons}
\label{sec:KKphotons}

So far we have discussed the WGC in the case where the relevant gauge fields originate from gauge fields in a higher-dimensional theory. Our findings supported the consistency of the WGC: the conjecture is stable under dimensional reduction and becomes monotonically weaker in the infrared when the radion is stabilized. This is reassuring. In this section we will consider the alternative case where a one-form gauge field in the dimensionally reduced theory contains an admixture of the KK photon arising from the graviton of the higher-dimensional theory. We will find, less reassuringly, that the WGC is potentially violated in the lower-dimensional theory even if it was satisfied in the higher-dimensional theory. Rather than undermining the WGC, we interpret this to mean that a stronger condition than the original WGC---discussed in \S\ref{sec:LWGC}---must be satisfied.

\subsection{The KK photon}

We begin by considering the case of a pure KK photon. The metric ansatz for reducing on a circle of radius $R$ with a radion $\lambda$ and a KK photon $B_1$ is
\be
ds^2 = e^{\frac{\lambda(x)}{d-2}} d{\hat s}^2(x) + e^{-\lambda(x)} (dy+R B_1)^2 \,,
\label{eq:ansatz}
\ee
where $y \cong y + 2 \pi R$ and $B_1$ is normalized so that the KK modes carry integral charges. The dimensionally reduced action is
\be
S = \frac{1}{2\kappa_d^2} \int d^d x \sqrt{-{\hat g}}\left[{\cal {\hat R}}_d - \frac{d-1}{4(d-2)} (\nabla \lambda)^2 - \frac{R^2}{2} e^{- \frac{d-1}{d-2} \lambda} H_2^2\right] \,,
\ee
where $H_2 = d B_1$. Thus, the KK photon gauge coupling and the radion--KK photon coupling are
\be
\frac{1}{e_{\rm KK}^2} = \frac{1}{2} R^2 M_d^{d-2} \;\;,\;\; \alpha_{\rm KK} = \sqrt{\frac{2(d-1)}{d-2}} \,,
\ee
where the latter is defined by the coupling to the normalized radion $\hat{\lambda} = \sqrt{\frac{d-1}{2(d-2)}} \lambda$.

From~(\ref{eq:generalWGC}), we find the WGC bound for the KK photon:
\be
\left[\frac{\alpha^2_{\rm KK}}{2} + \frac{d-3}{d-2}\right] m^2 \leq e_{\rm KK}^2 q^2 M_d^{d-2} \,.
\ee
Thus $\gamma_{\rm KK} = 1/2$ and
\be
m^2 \leq \frac{q^2}{R^2} \,.
\ee
Conversely, the KK spectrum of a particle with mass $m_0$ in the higher-dimensional theory is 
\be
m^2 = m_0^2 + \frac{q^2}{R^2} \,,
\ee
where the KK charge $q \in \mathbb{Z}$ specifies the quantized momentum $q/R$ along the compact circle. Thus, massless particles in the higher dimensional theory generate KK modes which saturate the WGC bound, whereas massive particles do not satisfy the bound. Since the higher-dimensional theory necessarily contains a massless graviton, the WGC bound for the KK photon is saturated. As usual, stabilizing the radion leads to a weaker WGC bound, which is then satisfied but not saturated.

We briefly consider magnetic charges. The Kaluza-Klein monopole \cite{Sorkin:1983ns, Gross:1983hb} is a smooth gravity solution in $d+1$ dimensions that appears as a magnetic monopole in the $d$-dimensional theory. For $d = 4$, the monopole mass is
\be
M_{\rm KK;mag} = \pi M_{\rm Pl}^2 R \,,
\ee
and its magnetic charge is ${\tilde Q} = 1$. Using the fact that $g_{\rm KK}^2 = 4\pi^2/e_{\rm KK}^2$, we see that this saturates the magnetic WGC for the KK photon:
\be
\gamma_{\rm KK} g_{\rm KK}^2 {\tilde Q}^2 = \kappa_d^2 M_{\rm KK;mag}^2.
\ee
The same is true for $N$-monopole states as well, which in the limit of coincident monopole positions may be thought of as a marginally bound single-particle state of charge ${\tilde Q} = N$ and mass $N M_{\rm KK;mag}$. For $d > 4$, the KK monopole is a brane with $d-3$ spacetime dimensions which is the product of the 4D KK monopole solution and additional flat spatial dimensions, so it continues to be true that the magnetic WGC bound is marginally satisfied.

\subsection{Black holes with both Kaluza-Klein charge and other $U(1)$ charge}
\label{subsec:blackholebothcharge}

The weak gravity conjecture for a combination of multiple $U(1)$s is more stringent than for each $U(1)$ individually: the set of charges must satisfy the ``convex hull condition''~\cite{Cheung:2014vva}. We have just seen that for a single KK photon in a theory with an unstabilized radion, the bound is satisfied only {\em marginally}. This suggests that there is no freedom to consider other photons in addition to the Kaluza-Klein photon. To test this intuition, we will explicitly construct black hole solutions that are charged under two $U(1)$ gauge groups, one of which is a Kaluza-Klein gauge group and one of which is arbitrary. The idea is as follows: we consider a gauge theory in $d$ dimensions that results from compactifying a $D = d+1$ dimensional gauge theory. We have already found solutions to the $d$-dimensional theory that are charged under the $U(1)$ inherited from the $D$-dimensional theory. To find solutions that have both charges, we first lift these solutions to $D$ dimensions by taking them to be constant in the extra dimension. The lifted solutions are charged black strings. Ordinarily we think of a black string as carrying charge under 2-form gauge fields, not 1-form gauge fields, but these instead carry 1-form charge that is smeared out along the string. Such solutions have been studied in the past \cite{Horowitz:2002ym}. Once we have this lifted black string solution, we can boost it in the extra dimension, then compactify back down to $d$ dimensions. The momentum acquired by the string through boosting becomes charge under the KK $U(1)$ in the lower-dimensional theory.

In addition, we can consider reduction of boosted black string solutions charged under 2-forms in the higher dimensional theory.  We will explore each of these cases in turn and see that they lead to rather distinct convex hull conditions.

\subsubsection{Dimensional reduction of black strings charged under 1-form gauge fields}\label{ssec:1formstring}

Beginning with the Einstein--Maxwell--dilaton theory in $D$ dimensions using the conventions of (\ref{eq:generalaction}), we reduce to $d = D-1$ dimensions with the ansatz:
\begin{align}
  d s^2_D &= e^{\frac{\lambda}{d - 2}} d \hat{s}_d^2 + e^{- \lambda}  (d y + R B_1)^2\,, \nonumber \\
  F_2^{(D)} &= \tilde{F}_2 + \frac{1}{2\pi} F_1 \wedge \left(\frac{d y}{R} + B_1\right) \,, & 
   A_1^{(D)} &= A_1 + \frac{A_0}{2\pi} \left(\frac{dy}{R}+B_1 \right)\,. \label{eq:dimredansatz2}
\end{align}
Here $A_0$ is a compact axion field with field strength $F_1 = d A_0$ and $A_0 \to A_0 + 2\pi$ is a gauge symmetry inherited from large gauge transformations around the circle in the $D$-dimensional theory. The field strength $\tilde{F}_2 = d A_1 + \frac{A_0}{2\pi} H_2$ satisfies the Bianchi identity
\be
  d \tilde{F}_2 = \frac{1}{2\pi} H_2 \wedge F_1 \,.
\ee
Such modified Bianchi identities are a familiar consequence of Kaluza-Klein theories (see, e.g., section 12.1 of \cite{Polchinski:1998rr}). The dimensionally reduced action is then:
\begin{align}
  S &= \frac{1}{2 \kappa_d^2}  \int d^d x \sqrt{- g}  \left[
  \mathcal{R}_d- \frac{d - 1}{4 (d - 2)}  (\nabla \lambda)^2 - \frac{1}{2} 
  (\nabla \phi)^2 \right] - \frac{1}{2} f^2 \int d^d x \sqrt{-g} e^{-\alpha \phi + \lambda} (\nabla A_0)^2 \nonumber \\
  &\mathrel{\phantom{=}} - \frac{1}{2 e_d^2}  \int d^d x \sqrt{- g}  e^{- \alpha \phi - \frac{\lambda}{d - 2}}
  | \tilde{F}_2 |^2 - \frac{1}{2 e_{KK}^2} \int d^d x \sqrt{-g} \, e^{- \frac{d - 1}{d - 2} \lambda}  | H_2 |^2, \label{eq:reducedaction}
\end{align}
where we introduce the axion decay constant $f = \frac{1}{2\pi R e_d}$.

We now consider a Lorentz boost in the $D$-dimensional theory 
\be
  t \rightarrow u t + v y \,, \qquad y \rightarrow u y + v t\,,
\ee
where $u^2 - v^2 = 1$. For simplicity, we only turn on certain parts of the full ansatz:\footnote{Here of course $g_{t t}<0$.}
\begin{align}
d\hat{s}_d^2 &= g_{t t} d t^2 + g_{i j} d x^i d x^j\,, &
{\tilde F}_2 &= {\tilde F}_{ti} dt \wedge dx^i\,, & F_1 &= F_i dx^i\,, & B_1 &= B_t dt,
\end{align}
omitting the components $g_{t i}$, $B_i$, $\tilde{F}_{i j}$ and $F_t$, as these are not present in the original background, nor are they generated by boosts.
With this ansatz, we find that the boosted fields are
\begin{align}
  e^{- \lambda'} &= e^{- \lambda}  (u + v R B_t)^2 + v^2 e^{\frac{\lambda}{d-2}} g_{t t}\,, & 
  e^{-\frac{d-3}{d-2} \lambda'} g'_{t t} &= e^{-\frac{d-3}{d-2} \lambda}  g_{t t}\,, \nonumber \\
  e^{- \lambda'} R B'_t &= e^{- \lambda}  (u + v R B_t) (u R B_t + v) + u v e^{\frac{\lambda}{d-2}} g_{t t}\,, &
  e^{\frac{\lambda'}{d-2}} g'_{i j} &= e^{\frac{\lambda}{d-2}} g_{i j}\,, \nonumber\\
  e^{- \lambda'}  \tilde{F}'_{t i} &= e^{- \lambda}  (u + v R B_t) \tilde{F}_{t i} + v e^{\frac{\lambda}{d - 2}}  g_{t t} \frac{F_i}{2\pi R}\,, & 
  \frac{F'_i}{2 \pi R} &=  (u + v R B_t) \frac{F_i}{2 \pi R} -  v \tilde{F}_{t i}\,,
\end{align}
in Einstein frame. In particular, the metric components are rescaled by a radion-dependent factor.

We consider a black hole that is electrically charged under $\tilde{F}_2$, but with $F_1 = H_2 = 0$. This is a special case of the discussion in~\S\ref{sec:dilatonicbranes}. Explicitly, we find the unboosted background
\begin{align}
  d s_d^2 &= - f_+ f_-^{\gamma_{\parallel}-1} d t^2 + f_+^{-1} f_-^{\gamma_{\perp} - 1} d r^2 + r^2 f_-^{\gamma_{\perp}} d \Omega_{d - 2}^2\,, &
  e^{\lambda} &= f_-^{\frac{2\gamma}{d - 1}}\,,\nonumber\\
  \tilde{F}_2 &= \frac{e_d}{\kappa_d} \frac{\cal Q}{r^{d - 2}} d t \wedge d r\,, & e^{\alpha \phi} &= f_-^{\alpha^2 \gamma} \,, \label{eqn:unboosted}
\end{align}
where $\gamma = \gamma_{1,d+1}(\alpha)$ from (\ref{eq:gammadef}) and 
\begin{align}
  \gamma_{\parallel} &=\frac{2 (d-3)}{d-2} \gamma\,, & \gamma_{\perp} &= \frac{2}{d - 3} - \frac{2\gamma}{d - 2}\,, &
  {\cal Q} &= (d - 3) \gamma^{\frac{1}{2}}  (r_+ r_-)^{\frac{d - 3}{2}}\,. \label{eq:usefulconstants}
\end{align}
Applying the boost to the radion, we obtain
\be
  e^{ - \lambda'} = e^{-\lambda} (u^2 - v^2 f_+ f_-^{2\gamma - 1}) \equiv e^{-\lambda} f_{\lambda}.
\ee
The fully boosted solution has a simple expression in terms of $f_\lambda$:
\begin{align}
  d s^2 &= - f_+ f_-^{\gamma_{\parallel}-1} f_{\lambda}^{-\frac{d - 3}{d - 2}} d t^2 + f_+^{- 1} f_-^{\gamma_{\perp} - 1} f_{\lambda}^{\frac{1}{d - 2}} d r^2 
  + r^2 f_-^{\gamma_{\perp}} f_{\lambda}^{\frac{1}{d - 2}} d \Omega_{d - 2}^2 \,, &  H_2 &= \frac{u}{v R}\, d t \wedge d f_{\lambda}^{- 1}\,, \nonumber \\
  \tilde{F}_2 &= \frac{e_d}{\kappa_d}\,\frac{u {\cal Q}}{r^{d - 2} f_{\lambda}} d t \wedge d r \,, & e^{\lambda} &= f_-^{\frac{2\gamma}{d - 1}} f_{\lambda}^{- 1} \,, \nonumber \\
  A_0 &= \theta +2\pi R\frac{ e_d}{\kappa_d} \frac{v {\cal Q}}{(d-3) r^{d - 3}} \,, & e^{\alpha \phi} &= f_-^{\alpha^2 \gamma} \,, \label{eqn:boostedsoln}
\end{align}
where we introduce the integration constant $\theta$ to allow for a background axion vev.

The ADM mass of this solution is
\begin{equation}
  M_{\tmop{ADM}} = \frac{V_{d-2}}{2 \kappa_d^2} \left[(d-2)  (r_+^{d-3} + (\gamma_{\parallel} - 1) r_-^{d-3}) + v^2 (d-3) (r_+^{d-3} + (2 \gamma - 1) r_-^{d-3}) \right] \,.
\end{equation}
To compute its charges, we first clarify how~(\ref{eq:Qel}) is modified due to the axion-induced coupling between $A_1$ and $B_1$. Electric charges are defined by their worldline actions
\be
S = \int_\Sigma (Q_F A_1 + Q_H  B_1) \,,
\ee
where $\Sigma$ is the worldline of a charged particle and $Q_F, Q_H$ must be integers for $e^{i S}$ to be invariant under large gauge transformations. The $A_1$ and $B_1$ equations of motion in the presence of these sources are
\begin{align}
d  \left[\frac{1}{e_d^2} e^{- \alpha \phi - \frac{\lambda}{d - 2}} \star\tilde{F}_2\right] &= Q_F \delta(\Sigma) \,, \\
d \left[ \frac{1}{e_{KK}^2} e^{- \frac{d - 1}{d - 2} \lambda} \star H_2 + \frac{A_0}{2 \pi e_d^2} e^{- \alpha \phi - \frac{\lambda}{d - 2}} \star \tilde{F}_2 \right] &= Q_H \delta(\Sigma) \,,
\end{align}
where $\delta(\Sigma)$ is a Dirac delta $(d-1)$-form corresponding to the worldvolume $\Sigma$. Applying Stokes' theorem, we conclude that
\begin{align} 
Q_F &= \frac{1}{e_d^2} \int_{S^{d-2}} e^{- \alpha \phi - \frac{\lambda}{d - 2}} \star\tilde{F}_2 \,, \label{eqn:QFinteg} \\
Q_H &= \frac{1}{e_{KK}^2} \int_{S^{d-2}} e^{- \frac{d - 1}{d - 2} \lambda} \star H_2 + \frac{1}{e_d^2} \int_{S^{d-2}} \frac{A_0}{2\pi}\, e^{- \alpha \phi - \frac{\lambda}{d - 2}} \star \tilde{F}_2 \,, \label{eqn:QHinteg}
\end{align}
which differs from~(\ref{eq:Qel}) for $A_0 \ne 0$. Notice that $Q_H \to Q_H + n Q_F$ for $A_0 \to A_0 + 2 \pi n$. This corresponds to the fact that $A_1 \to A_1 - n B_1$ under the same large gauge transformation. $Q_H$ is an example of a Page charge~\cite{Page:1984qv, Marolf:2000cb}, which is conserved and quantized, but not invariant under large gauge transformations. Using~(\ref{eqn:QFinteg}, \ref{eqn:QHinteg}), we obtain
\begin{align}
Q_F &= \frac{V_{d-2}}{e_d \kappa_d} u \mathcal{Q}\,, \\
Q_H &= \frac{R V_{d-2}}{2 \kappa_d^2} u v \mathcal{Q}_H+ \frac{\theta}{2\pi} Q_F\,, & \Big(\mathcal{Q}_H &\equiv (d - 3)  [r_+^{d - 3} + (2 \gamma - 1) r_-^{d - 3}]\Big) \,.
\end{align}

To interpret the boosted solution~(\ref{eqn:boostedsoln}), we analyze the behavior of $f_{\lambda} = u^2 - v^2 (f_+/f_-) f_-^{2 \gamma}$,
which depends on $\gamma>0$. For $r > r_+ \geqslant r_- > 0$, $f_+/f_-$ decreases monotonically from $1$ to $0$ as $r$ decreases to $r_+$. Since $f_-$ is also monotonically decreasing on the same interval, we conclude that $f_\lambda$ increases monotonically from $1$ to $u^2$ as $r$ decreases to $r_+$. For $r<r_+$, the behavior is $\gamma$-dependent. One can show that for $\gamma>1/2$, $f_\lambda$ increases to a finite maximum before decreasing back to $u^2$ at $r=r_-$, whereas for $\gamma<1/2$, $f_\lambda$ increases monotonically, diverging at $r=r_-$. For $\gamma=1/2$, $f_\lambda$ increases monotonically but reaches a finite maximum $f_\lambda = 1 + v^2 \frac{r_+^{d-3}}{r_-^{d-3}}$ at $r=r_-$.

\begin{figure}
\begin{center}
\includegraphics[width=0.55\textwidth]{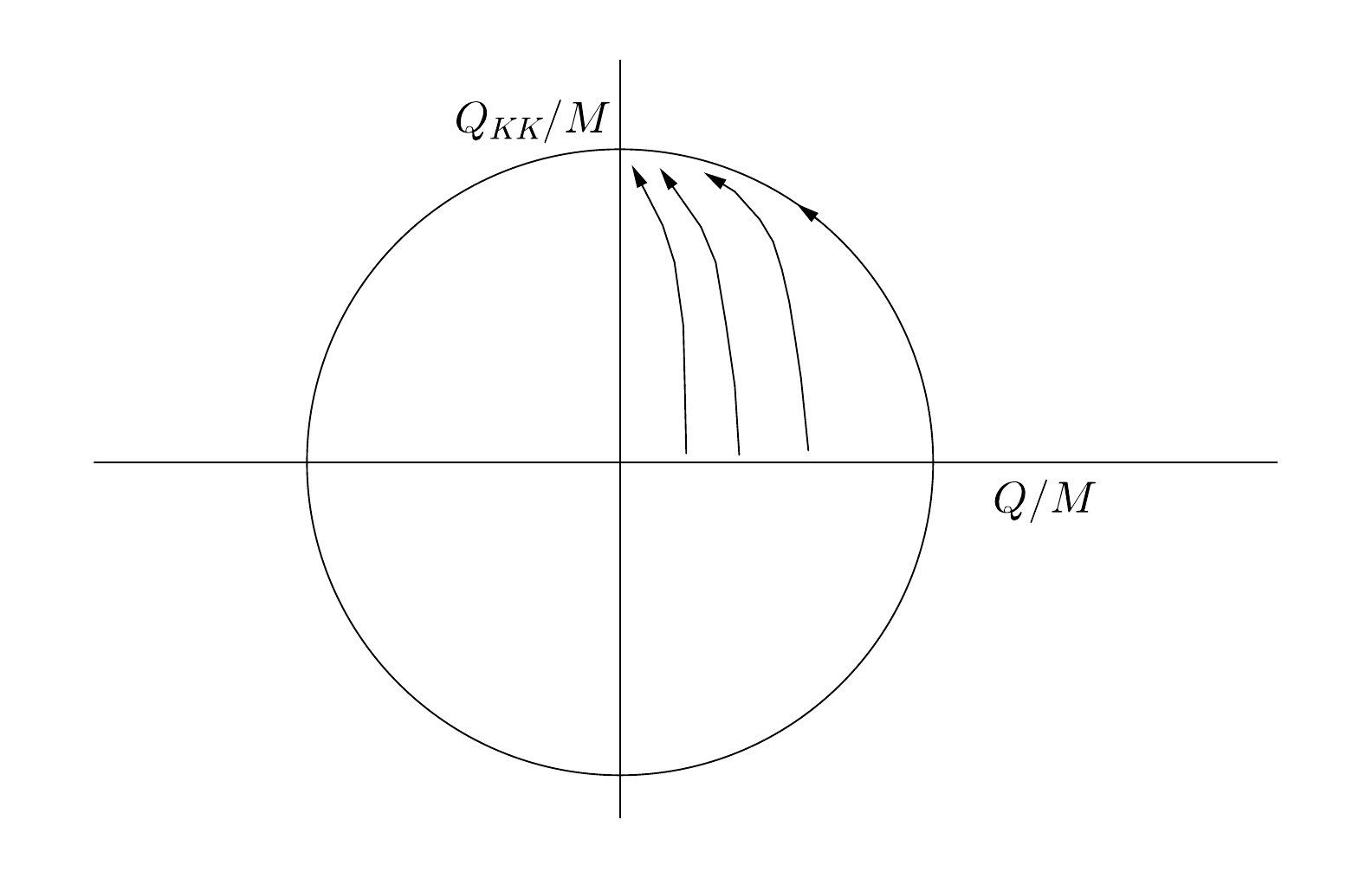}
\caption{Boosting a black string charged under a 1-form. Finite boosts of sub-extremal black strings remain sub-extremal black holes after dimensional reduction. Infinite boosts yield the extremal KK charged black hole, whereas finite boosts of extremal black strings map out the remaining extremal black holes with both charges nonzero.\label{fig:boostedBH}}
\label{boosted}
\end{center}
\end{figure}

In each case, $f_{\lambda}$ is finite and positive for $r > r_-$. This implies
in particular that there is a smooth event horizon at $r = r_+$ for any $r_- < r_+$.\footnote{For $r_+ = r_-$, the dilaton
$\phi$ blows up at the horizon---just as in the unboosted case---regardless of whether $f_{\lambda}$ is finite or not.} These black holes are
therefore sub-extremal. To obtain an extremal black hole from the boost, we
must either start with an extremal black hole ($r_+ = r_-$), or perform an infinite boost. In the latter case, the ADM mass for $u \gg 1$ is
\begin{equation}
  M \longrightarrow \frac{V_{d - 2}}{2 \kappa_d^2}  (d - 3) u^2 [
  r_+^{d - 3} + (2 \gamma - 1) r_-^{d - 3}] \equiv \frac{V_{d - 2}}{2 \kappa_d^2}  (d - 3) r_0^2 \,.
\end{equation}
so we must take $r_{\pm} \to 0$ at the same time to hold the mass (hence $r_0$) fixed. In this limit $Q_F \to 0$ but $Q_H \to \frac{R V_{d-2}}{2 \kappa_d^2} (d-3) r_0^2$, and we recover an extremal KK charged black hole ($M = Q_H/R$), where $f_{\pm} \to 1$ and $f_{\lambda} \to r_0^{d-3}/r^{d-3}$. Thus, infinite boosts converge on the extremal KK-charged black hole regardless of $r_+, r_-$, whereas finite boosts of extremal $\tilde{F}_2$-charged black holes remain extremal. This situation is illustrated in Figure~\ref{fig:boostedBH}.

To determine the extremality bound, we set $r_+ = r_-$ in the boosted solution, giving
\begin{align}
M &= \frac{V_{d - 2}}{\kappa_d^2} u^2 (d - 3) \gamma r_+^{d - 3} \,, \nonumber \\
Q_F &= \frac{V_{d-2}}{e_d \kappa_d} u (d - 3) \gamma^{\frac{1}{2}}  r_+^{d-3}\,, &
Q_H &= \frac{R V_{d-2}}{\kappa_d^2} u v (d - 3) \gamma r_+^{d - 3}+ \frac{\theta}{2\pi} Q_F\,.
\end{align}
We observe that:
\begin{equation}
  M^2 = \gamma e_d^2 M_d^{d - 2} Q_F^2 + \frac{1}{R^2}  \left(
  Q_H - \frac{\theta}{2 \pi} Q_F \right)^2 \,.
\end{equation}
It is straightforward to check that sub-extremal black holes with the same charges have a larger mass, hence
\begin{equation} \label{eqn:KKbound1}
  M^2 \ge \gamma e_d^2 M_d^{d - 2} Q_F^2 + \frac{1}{R^2}  \left(
  Q_H - \frac{\theta}{2 \pi} Q_F \right)^2 \,,
\end{equation}
and the inverted inequality sets the appropriate weak gravity constraint on the particle spectrum.

\subsubsection{Dimensional reduction of black strings charged under 2-form gauge fields}\label{ssec:2formstring}

We now consider black holes charged under the KK photon as well as a photon descending from a two-form in $D$ dimensions. This case is closely analogous to that discussed in the previous section, so we will be brief, highlighting the differences. We reduce the two-form with the ansatz:
\begin{align}
F_3^{(D)} &= \tilde{F}_3 + \frac{1}{2\pi} F_2 \wedge \left(\frac{d y}{R} + B_1\right) \,, & 
   A_2^{(D)} &= A_2 + \frac{A_1}{2\pi} \left(\frac{dy}{R}+B_1 \right)\,. 
\end{align}
Since we are interested in black holes in $d$ dimensions, we can set $\tilde{F}_3 = A_2 = 0$. The dimensionally-reduced action for the remaining fields is
\begin{align}
  S &= \frac{1}{2 \kappa_d^2}  \int d^d x \sqrt{- g}  \left[
  \mathcal{R}_d- \frac{d - 1}{4 (d - 2)}  (\nabla \lambda)^2 - \frac{1}{2} 
  (\nabla \phi)^2 \right]  \nonumber \\
  &\mathrel{\phantom{=}} - \frac{1}{2 e_d^2}  \int d^d x \sqrt{- g}  e^{- \alpha \phi + \frac{d-3}{d - 2}\lambda}
  | F_2 |^2 - \frac{1}{2 e_{KK}^2} \int d^d x \sqrt{-g} \, e^{- \frac{d - 1}{d - 2} \lambda}  | H_2 |^2, \label{eq:reducedaction2}
\end{align}
which is similar to~(\ref{eq:reducedaction2}) except that $\lambda$ couples differently to $F_2$ and there is no axion. We then consider a black hole with vanishing KK charge, of the form~(\ref{eqn:unboosted}), except that now
\be
e^{-\lambda} = f_-^{\frac{2 (d-3)}{d-2} \gamma} \,.
\ee
Boosting, we obtain:
\be
e^{-\lambda'} = e^{-\lambda} \left(u^2 - v^2 f_+/f_- \right) \equiv e^{-\lambda} \tilde{f}_\lambda \,.
\ee
The full boosted background admits a simple expression in terms of $\tilde{f}_\lambda$ (cf.~(\ref{eqn:boostedsoln})):
\begin{align}
  d s^2 &= - f_+ f_-^{\gamma_{\parallel}-1} \tilde{f}_{\lambda}^{-\frac{d - 3}{d - 2}} d t^2 + f_+^{- 1} f_-^{\gamma_{\perp} - 1} \tilde{f}_{\lambda}^{\frac{1}{d - 2}} d r^2 
  + r^2 f_-^{\gamma_{\perp}} \tilde{f}_{\lambda}^{\frac{1}{d - 2}} d \Omega_{d - 2}^2 \,, \span\omit\span &  e^{-\lambda} &= f_-^{\frac{2 (d-3)}{d-2} \gamma} \tilde{f}_{\lambda} \,, \nonumber \\
  F_2 &= \frac{e_d}{\kappa_d}\,\frac{{\cal Q}}{r^{d - 2}} d t \wedge d r \,, &  H_2 &= \frac{u}{v R}\, d t \wedge d \tilde{f}_{\lambda}^{- 1}\,, & e^{\alpha \phi} &= f_-^{\alpha^2 \gamma} \,, 
  \label{eqn:boostedsoln2}
\end{align}
where $\mathcal{Q}, \gamma_{\perp}, \gamma_{\parallel}$ are given by~(\ref{eq:usefulconstants}) and $\gamma = \gamma_{2;d+1}(\alpha)$. Note that $F_2$ is unaffected by the boost---unlike before---because $F_3 \propto dt \wedge dy$ is boost-invariant. 
Computing the ADM mass and charge, we obtain:
\begin{align}
M_{\rm ADM} &= \frac{V_{d-2}}{2 \kappa_d^2} \left[(d-2)  (r_+^{d-3} + (\gamma_{\parallel} - 1) r_-^{d-3}) + v^2 (d-3) (r_+^{d-3} - r_-^{d-3}) \right] \,,\span\omit\span\omit \nonumber \\
Q_F &= \frac{V_{d-2}}{e_d \kappa_d}   \gamma^{\frac{1}{2}} (d - 3)  (r_+ r_-)^{\frac{d - 3}{2}}\,, 
& Q_H &= \frac{R V_{d-2}}{2 \kappa_d^2} u v (d - 3)  (r_+^{d - 3} - r_-^{d - 3})\,.
\end{align}

\begin{figure}
\begin{center}
\includegraphics[trim=25mm 30mm 15mm 0mm, clip, width=0.50\textwidth]{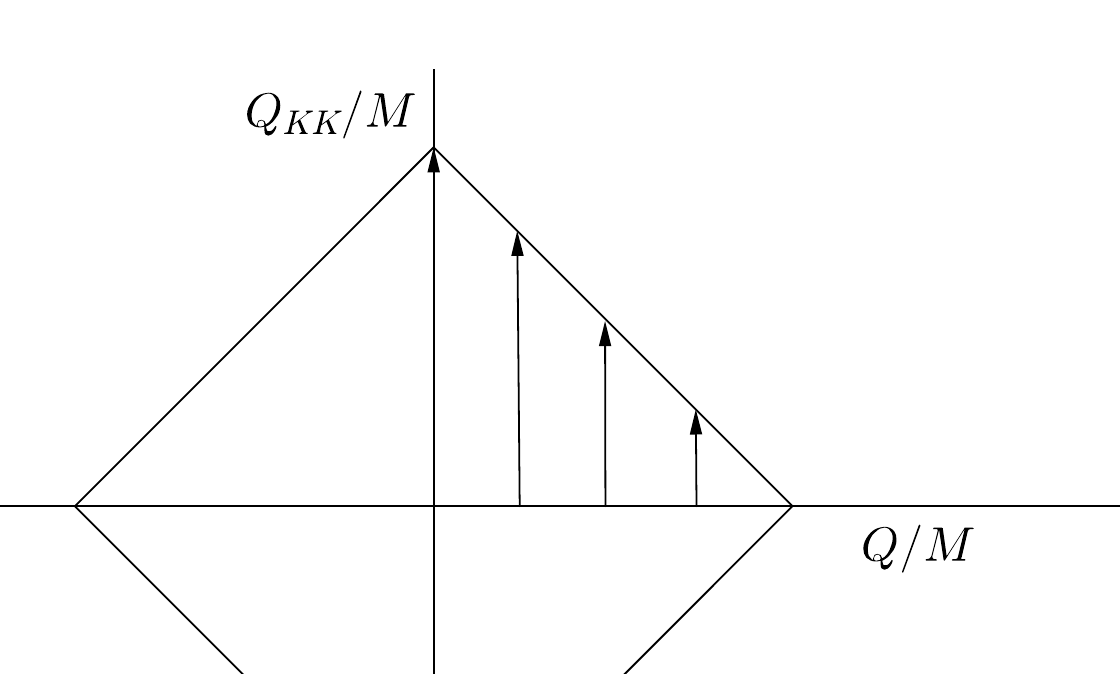}
\caption{Boosting a black string charged under a 2-form.  Here, the extremal black string is invariant under boosts, and instead extremal black holes in the dimensionally reduced theory come from maximally boosting while simultaneously taking $r_+ \rightarrow r_-$.  Interestingly, the extremality condition takes the form of a linear relation $M \geq c_F |Q_F| + c_H |Q_H|$ rather than a quadratic one.}
\label{squarehull}
\end{center}
\end{figure}

As before, (\ref{eqn:boostedsoln2}) has a smooth horizon for $r_+ > r_-$ and $u,v$ finite, so extremal black holes require $r_+ = r_-$ and/or an infinite boost. However, unlike before, the case $r_+ = r_-$ is boost invariant (corresponding to a relativistic black string in $D+1$ dimensions). Since taking $u \to \infty$ with $r_+/r_-$ fixed sends $Q_F/M \to 0$, to obtain extremal black holes charged under both $F_2$ and $H_2$, we must simultaneously take $u \to \infty$ and $r_- \to r_+$. To do so, we hold fixed the combinations
\begin{align}
r_F^{d-3} &\equiv (r_+ r_-)^{\frac{d - 3}{2}}\,, & r_H^{d-3} &\equiv u v (r_+^{d - 3} - r_-^{d - 3}) \,,
\end{align}
while taking $u,v \to \infty$. This gives
\begin{align}
M_{\rm ADM} &= \frac{V_{d-2}}{2 \kappa_d^2} (d-3) (2 \gamma r_F^{d-3}  + r_H^{d-3} ) \,, \nonumber \\
Q_F &= \frac{V_{d-2}}{e_d \kappa_d}   \gamma^{\frac{1}{2}} (d - 3) r_F^{d-3} \,, 
& Q_H &= \frac{R V_{d-2}}{2 \kappa_d^2} (d - 3) r_H^{d-3} .
\end{align}
Different signs can be obtained for $Q_F$ and $Q_H$ by starting with a black hole of opposite charge and/or by boosting $u \to \infty, v \to -\infty$. Thus, the extremality bound is
\be \label{eqn:diamondhull}
\kappa_d M \geq \sqrt{\gamma} e_d \left|Q_F\right| + \sqrt{\gamma_{\rm KK}} e_{\rm KK} \left|Q_H\right|.
\ee
As expected, this reduces to the earlier extremality bounds when either $Q_F$ or $Q_H$ is zero, but it interpolates between them linearly rather than quadratically. This is shown in Figure \ref{squarehull}.

Extremality bounds of the form~(\ref{eqn:diamondhull}) occur in string theory as BPS bounds in cases where objects charged under the two $U(1)$s are mutually BPS. For instance, exactly the situation described in this section occurs in type IIA string theory compactified on a circle. D0 branes and F-strings wrapped around the compact circle are mutually BPS, hence combinations of $M$ D0 branes and $N$ wrapped strings combine with zero binding energy and saturate a bound of the form~(\ref{eqn:diamondhull}). By contrast, BPS bounds of the form~(\ref{eqn:KKbound1}) occur when charged objects which are not mutually BPS bind together with non-zero binding energy to saturate a BPS bound, such as $(p,q)$ strings in type IIB string theory, which are bound states of $p$ F-strings and $q$ D1 branes.

\subsection{The convex hull condition}
\label{subsec:convexhull}

We now consider the kinematics of black hole decay for the subextremal KK-charged black holes described in~\S\ref{subsec:blackholebothcharge}. As usual, for black holes to be able to decay to particles, the convex hull of the charge-to-mass ratios $\vec{\zeta} \equiv \vec{Q}/m$ of these particles should contain the charge-to-mass ratios of all possible subextremal black holes. Ordinarily, the latter form an ellipsoid centered on the origin of $\vec{\zeta}$-space---the ``black hole region,'' enclosed by an ``extremal boundary''---which can be viewed as a ball of unit radius in $\vec{Z}$-space, where $Z^a = L^a_{\; b} \zeta^b$ for some appropriate choice of $L^a_{\; b}$.
 However, the example described in~\S\ref{ssec:2formstring} illustrates that the black hole region may take a different shape when even one massless scalar is present, whereas that described in~\S\ref{ssec:1formstring} illustrates that the region sometimes remains ellipsoidal even with complicated scalar couplings.

We first consider the mixing between the KK photon and a photon descending from a two-form in $D$ dimensions, as in~\S\ref{ssec:2formstring}. We define the $\vec{Z}$-vector:
\be
\vec{Z} \equiv \frac{1}{m}\! \left(e_d\, M_d^{\frac{d-2}{2}} \gamma^{\frac{1}{2}}\, Q_F, \frac{Q_{\rm KK}}{R} \right) \,,
\ee
for a particle with mass $m$ and charges $Q_F$ and $Q_{\rm KK}$ under the two photons, so that the black hole region $|Z_F| + |Z_{\rm KK}| < 1$ is a diamond with corners at $(\pm 1,0)$ and $(0,\pm1)$. If the WGC is satisfied in $D$ dimensions, then there must be a string with charge $q$ and tension $T_0$ in the $D$-dimensional theory such that $Z_0 \equiv e_D M_D^{\frac{D-2}{2}} \gamma^{\frac{1}{2}} \frac{|Q_F|}{T_0} \ge 1$. This implies that there is a particle in the $d$-dimensional theory with mass $m = (2 \pi R) T_0$ and charge $Q_F=q$, coming from the string wrapped once around the compact circle. This particle has $\vec{Z}_{\rm string} = (\pm Z_0, 0)$, which lies outside the black hole region by assumption. Similarly, there are KK-charged particles, such as the KK modes of the graviton, with $\vec{Z}_{\rm KK} = (0,\pm 1)$, lying on the extremal boundary. The convex hull of these two particles covers the entire black hole region for $Z_0 \ge 1$, hence the WGC in $D$ dimensions implies the WGC in $d$ dimensions, as in all examples considered up to this point.

\begin{figure}
\begin{center}
\includegraphics[trim=25mm 25mm 30mm 0mm,clip,height=0.28\textheight]{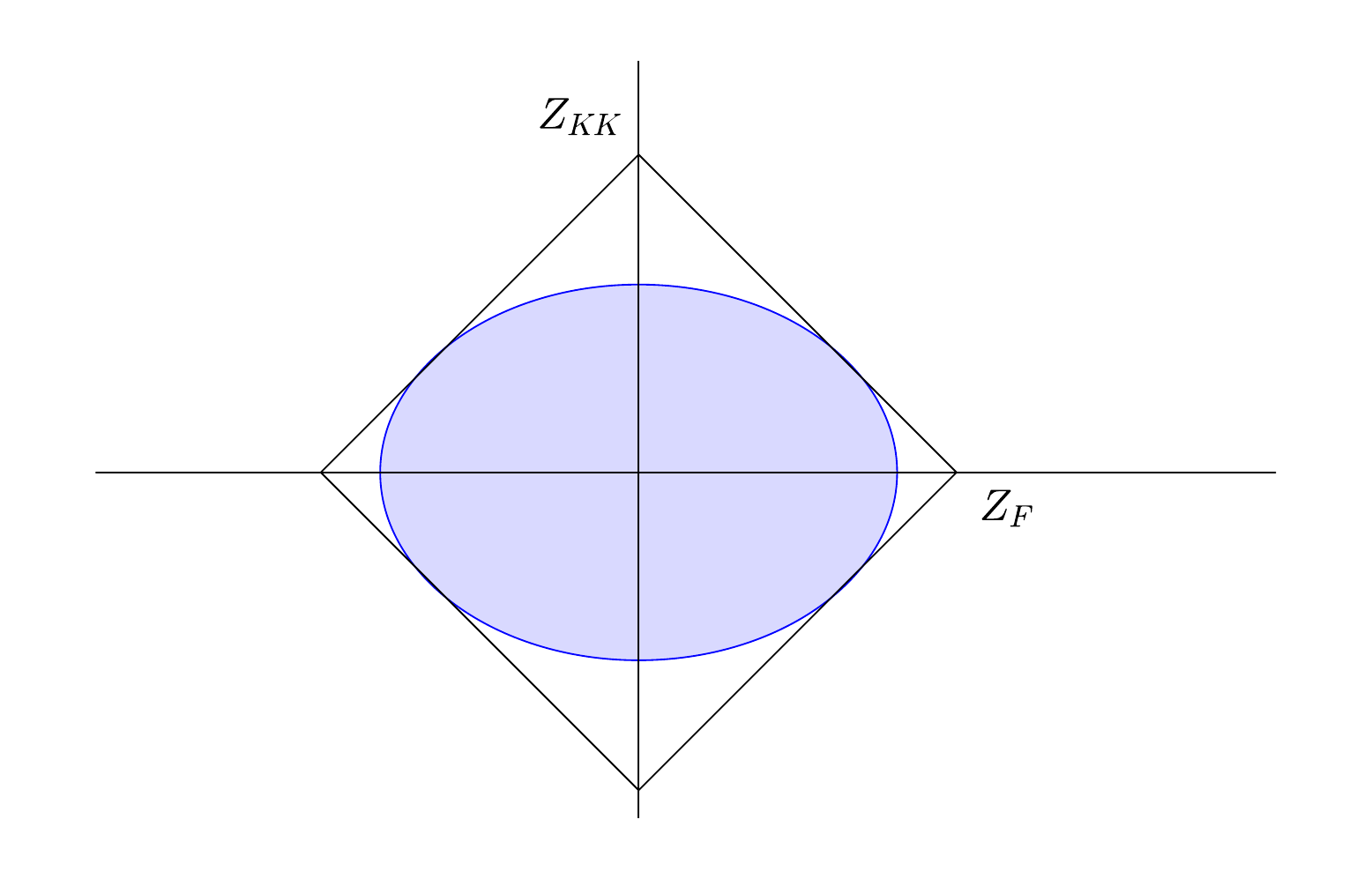}
\caption{Stabilizing the scalars shrinks the black hole region, which becomes ellipsoidal. In the case without a $D$-dimensional dilaton ($\alpha = 0$) the stabilized extremal boundary intersects the unstabilized extremal boundary at four points, whereas with a dilaton ($\alpha>0$), the stabilized extremal boundary lies strictly inside the unstabilized one.}
\label{stabilizedhull}
\end{center}
\end{figure}

Notice that this argument depends on the fact that the black hole region is diamond-shaped, rather than circular, since the KK modes lie on the extremal boundary. We might become concerned that, after stabilizing the radion and/or the dilaton, the circular black hole region that results will lead to a violation of weak gravity. In fact, in this as in every example in this paper, stabilization always shrinks the black hole region. After stabilization, the black hole region is the ellipsoid
\be
\frac{\gamma_0}{\gamma} Z_F^2 + \frac{\gamma_0}{\gamma_{\rm KK}} Z_{\rm KK}^2 = \left(\frac{\alpha^2}{2}\frac{d-2}{d-3} + \frac{2 (d-2)}{d-1}\right) Z_F^2 +\frac{2(d-2)}{d-3} Z_{\rm KK}^2 < 1 \,,
\ee
where $\gamma_0 \equiv \frac{d-2}{d-3}$ sets the stabilized extremality bound in $d$ dimensions and $\alpha$ is the dilaton coupling in $D$ dimensions. The unstabilized extremal boundary, at $|Z_{\rm KK}| + |Z_F| = 1$ minimizes $\frac{\gamma_0}{\gamma} Z_F^2 + \frac{\gamma_0}{\gamma_{\rm KK}} Z_{\rm KK}^2$ at
\be
|Z_F| = \left(\frac{\alpha^2}{4} + \frac{2 (d-2)}{d-1}\right)^{-1} = 1 - |Z_{\rm KK}| \,,
\ee
where
\be
\left[\frac{\gamma_0}{\gamma} Z_F^2 + \frac{\gamma_0}{\gamma_{\rm KK}} Z_{\rm KK}^2\right]_{\rm min} = 1+\frac{d-1}{d-3} \frac{\alpha^2}{4} |Z_F| \ge 1 \,.
\ee
Thus, the stabilized black hole region lies entirely within the unstabilized one, with their extremal boundaries touching at four points for $\alpha = 0$ and otherwise not intersecting. This situation is illustrated in Figure~\ref{stabilizedhull}. In part because the KK modes no longer touch the extremal boundary, the convex hull condition is satisfied.

We now consider the mixing between the KK photon and a photon reduced from a $D$-dimensional photon, as in~\S\ref{ssec:1formstring}. We define the $\vec{Z}$-vector:
\be
\vec{Z} \equiv \frac{1}{m}\! \left(e_d\, M_d^{\frac{d - 2}{2}} \gamma^{\frac{1}{2}}\, Q_F, \frac{1}{R} \left[Q_{\rm KK}-\frac{\theta}{2 \pi} Q_F\right]\right) \,,
\ee
for a particle with mass $m$ and charges $Q_F$ and $Q_{\rm KK}$ under the two photons, where $\theta$ is the vev of the axion. The black hole region is then the unit disk, $Z_F^2 + Z_{\rm KK}^2 < 1$. The WGC in $D$ dimensions implies that there is a particle of charge $Q_F=q$ and mass $m_0$ such that $Z_0 \equiv e_D M_D^{\frac{D-2}{2}} \gamma^{\frac{1}{2}} \frac{|q|}{m_0} \ge 1$. Dimensionally reducing, we obtain a tower of KK modes, with masses $m^2 = m_0^2 + \frac{1}{R^2} \big(n-\frac{q \theta}{2 \pi}\big)^2$ and charges $Q_F=q$ and $Q_{\rm KK} = n$, such that
\be
\vec{Z}_{(n)} = \frac{\left(\mu Z_0, x_n \right)}{\sqrt{\mu^2 + x_n^2}}  \,, \qquad \mu \equiv m_0 R\,, \qquad x_n \equiv n - \frac{q\theta}{2 \pi} \,.
\ee
The vectors $\vec{Z}_{(n)}$ lies on the ellipsoid $Z_F^2 / Z_0^2 + Z_{\rm KK}^2 =1$, outside the unit disk, so that each KK mode has sufficient charge to discharge a subextremal black hole with a proportional charge vector.

\begin{figure}
\begin{center}
\includegraphics[trim=45mm 40mm 0mm 0mm,clip,width=0.49\textwidth]{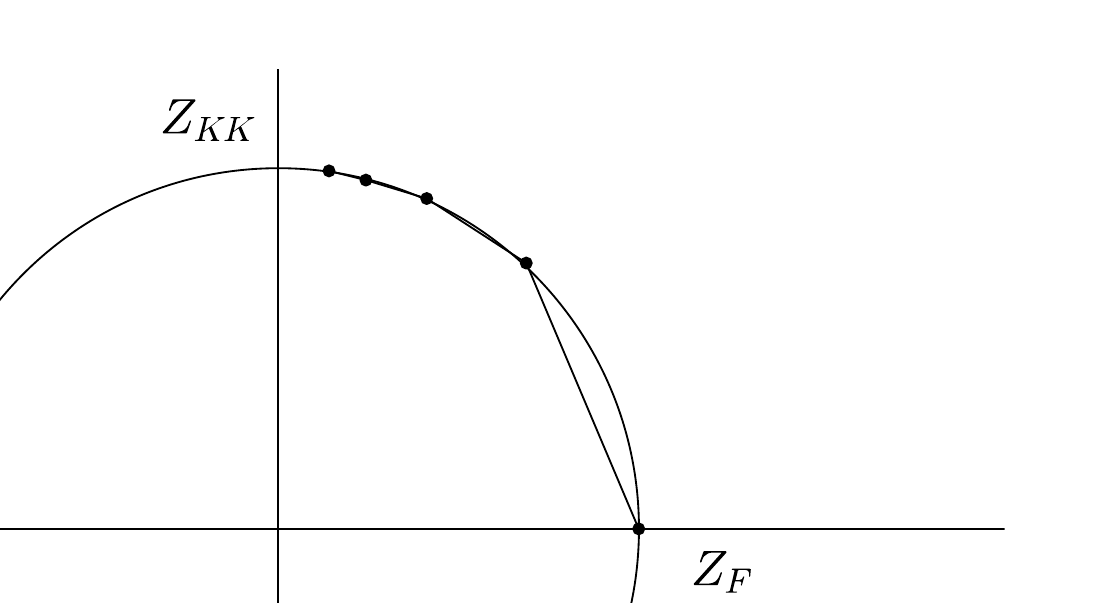}
\includegraphics[trim=43mm 40mm 0mm 0mm,clip,width=0.49\textwidth]{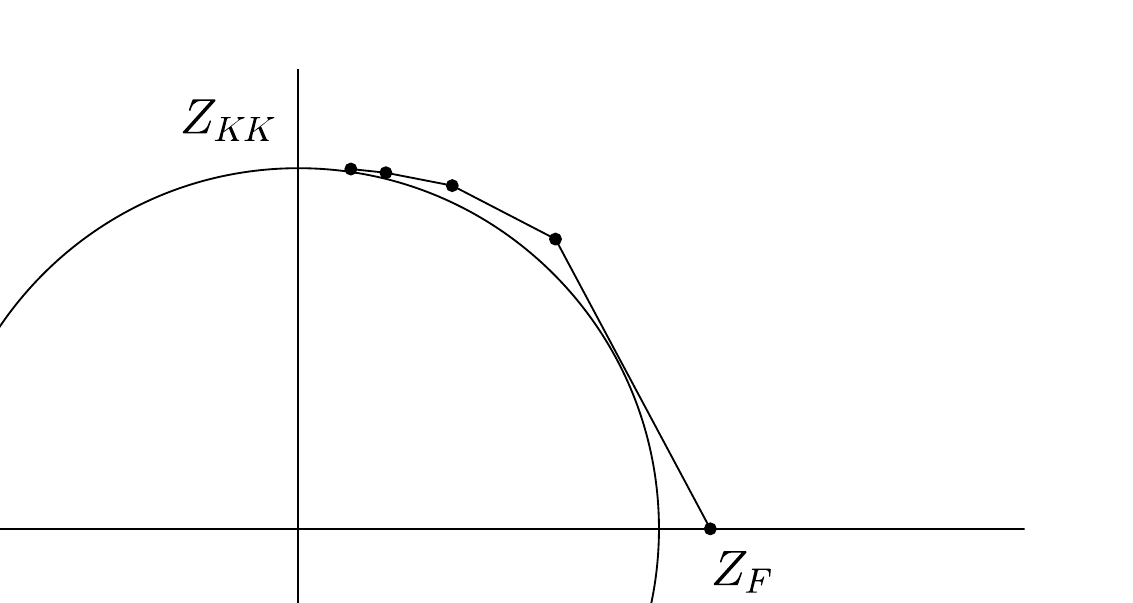}
\caption{The CHC for a theory with a KK $U(1)$ plus another $U(1)$.  It is possible for the charge-to-mass vector of every KK mode to obey $|\vec{Z}_n| \geq 1$ without satisfying the CHC, as shown at left.  Instead, the charge-to-mass vectors must be sufficiently large that the line segments connecting them lie outside the unit disk, as shown at right.}
\label{convexhull}
\end{center}
\end{figure}

However, this is not sufficient to ensure that the convex hull condition is satisfied, because the KK modes only populate a finite density of points along the ellipsoid $Z_F^2 / Z_0^2 + Z_{\rm KK}^2 =1$ away from from the poles, $(0,\pm1)$, and the lines between consecutive points can intersect the unit disk, as illustrated in Figure~\ref{convexhull}. In particular, the orthogonal distance to the origin of the line between $\vec{Z}_{(n)}$ and $\vec{Z}_{(n+1)}$ is
\be
\left[1+\frac{1}{Z_0^2} - \left(1+\frac{x_n x_{n+1}}{Z_0^2 \mu^2} \right)\left(1- \left(\sqrt{x_{n+1}^2 + \mu^2}-\sqrt{x_n^2 + \mu^2}\right)^2 \right) \right]^{-1/2} \,.
\ee
For a fixed value of $\theta$, the convex hull condition demands that this distance is at least one for all $n$. The strongest constraint comes from $-1\le x_n \le 0$ (when $x_n = - \left(\frac{q\theta}{2 \pi} \bmod 1\right)$), so that
\be
(m_0 R)^2 \ge \frac{1}{4 Z_0^2 (Z_0^2 - 1)} + \frac{n_0 (1-n_0)}{Z_0^2} \,, \qquad n_0 \equiv \frac{q\theta}{2 \pi} \bmod 1 \,.
\ee
Thus, for any value of $Z_0$, there is some minimum radius $R_{\rm min}$ below which the convex hull condition is not satisfied! Since we have assumed an unstabilized radion (e.g.\ due to unbroken supersymmetry), the radius is a modulus, and we should demand that the convex hull condition is satisfied everywhere in moduli space.

There are two ways to resolve this issue. Firstly, the effective field theory description we have been using will break down for $R \simlt \Lambda_{D}^{-1}$, where $\Lambda_D$ is the cutoff of the $D$-dimensional parent effective field theory. If $\Lambda_D \simlt R_{\rm min}^{-1}$, then the convex hull condition is satisfied everywhere in moduli space where the above calculation is under control, and there is no conflict with weak gravity. Alternatively, the $D$-dimensional theory may have a high cutoff but incorporate additional charged particles besides the one considered above. To solve the problem, these particles must also satisfy $Z \ge 1$. If there are two particles with the same charge, then only the lighter particle will contribute to the convex hull. Thus, the extra particles which contribute to the convex hull will all have distinct charges. It's easy to see that for any finite number of such particles, there is still a minimum radius $R_{\rm min}$, below which the convex hull condition is not satisfied. Therefore, to satisfy the convex hull condition everywhere in moduli space, we would need an infinite number of charges particles---each with a distinct charge---all satisfying $Z \ge 1$! 

In fact, these two explanations are related. A $D$-dimensional gauge theory coupled to an infinite tower of charged particles is generically badly behaved in the ultraviolet, signaling the need for a cutoff or some new organizing principle, such as a string theory or a theory with further extra dimensions. Conversely, in many examples the cutoff $\Lambda_D$ signals the appearance of new particles, such as the massive string states at the string scale, KK modes at the compactification scale, or composites at a confinement scale. These new particles must satisfy further constraints to ensure that black hole decay is kinematically allowed and will typically need to be charged, creating a similar situation to that analyzed above.

To illustrate how the convex hull condition is satisfied in a concrete example, let us suppose that the $D$-dimensional photon is itself the KK photon of a $(D+1)$-dimensional theory, so that the $d$-dimensional theory is equivalent to the $(D+1)$-dimensional theory compactified on a two-torus. The KK mass spectrum is
\be \label{eqn:torusmass}
m^2 = m_0^2 + \frac{(q_1 - \theta q_2)^2}{R_1^2}  + \frac{q_2^2}{R_2^2} \,,
\ee
where $q_1$ and $q_2$ denote the KK charges, $R_1$ and $R_2$ the radii of the corresponding circles, $m_0$ the mass of the particle in $D+1$ dimensions, and $\theta$ the associated axion. Notice that this takes the same form as~(\ref{eqn:KKbound1}). In fact, it's easy to check that the KK modes lie within the black hole region for $m_0 > 0$ and on the extremal boundary for $m_0 = 0$. Thus, the KK modes of the graviton (or another massless field) densely cover the extremal boundary at every angle of the form $p \pi/ q$ for $p, q \in \mathbb{Z}$ (covering every extremal black hole with quantized charges), and the convex hull covers the entire black hole region. By contrast, if we had truncated the $D$-dimensional KK spectrum to some finite number of modes before compactifying the second circle, then the extremal boundary would not be densely covered, and the convex hull condition would be violated, regardless of the compactification radius! Thus, the entire KK spectrum is needed in order to satisfy the WGC.

The preceding example occurs frequently in string theory. For instance, M-theory on a torus is dual to type IIB string theory on a circle, where the KK modes of the graviton are dual to $(p,q)$ strings wound around the circle. In this case, the KK modes are BPS states, and the mass formula~(\ref{eqn:torusmass}) is exact.

It's interesting to compare the above discussion to the case where the radion and dilaton are stabilized after compactification. As a result of stabilization, the black hole region lies entirely within the ellipse $Z_F^2 / Z_0^2 + Z_{\rm KK}^2 =1$ on which the KK modes appear. Since the axion is generically stabilized by non-perturbative effects at $\theta = 0$, for $Z_0 \simgt \sqrt{2}$ (depending somewhat on $\alpha$) the convex hull condition is satisfied for any stabilized radius $R$. However, if we also impose the WGC at local maxima of the scalar potential---the XWGC~\cite{Heidenreich:2015wga}---then putting $\theta= \pi/q$ will violate the convex hull condition if $R$ is too small, regardless of $Z_0$, in the absence of additional charged particles.

\section{The Lattice Weak Gravity Conjecture}
\label{sec:LWGC}

For the first time in this paper, we have encountered an example where the WGC in $D$ dimensions does not necessarily imply the WGC in $d$ dimensions. This is an appealing property of our earlier examples, and it is tempting to postulate a stronger form of the WGC that is always preserved under dimensional reduction. Given the discussion of the convex hull condition in~\S\ref{subsec:convexhull}, an obvious candidate is the following:\\

\noindent
{\bf The Lattice Weak Gravity Conjecture (LWGC):} For every point $\vec{Q}$ on the charge lattice, there is a particle of charge $\vec{Q}$ with charge-to-mass ratio at least as large as that of
a large, semi-classical, non-rotating extremal black hole with charge $\vec{Q}_{\rm BH} \propto \vec{Q}$.\\

\noindent
Note that this condition implies the convex hull condition, and is preserved under toroidal compactification, at least in the examples discussed in this paper.
  Moreover, there are certainly examples where the LWGC is true, such as for the KK reduction on a two-torus and cases related to this by string dualities. Although the LWGC implies an infinite number of charged particles, many of these particles will have super-Planckian masses, and can be interpreted as extremal black holes. For this to succeed, it is important that Planckian corrections to the black hole extremality bound reduce the mass of extremal black holes. Fortunately, there is some evidence for this hypothesis~\cite{Kats:2006xp}.

The LWGC is a strengthened form of the WGC which has not, to our knowledge, been previously discussed in the literature. Possible strong forms of the WGC were discussed already in \cite{ArkaniHamed:2006dz} and recently their possible importance both for understanding the WGC itself and for applications to inflation was stressed by \cite{Brown:2015iha}. The latter paper suggested a strong form of the WGC that requires that the lightest state (possibly consisting of multiple particles) in any {\em direction} in charge space be superextremal, which would follow from the LWGC. We require a superextremal state for all {\em points} in the charge lattice, rather than all directions, because we have seen that the WGC can fail after compactification otherwise. Single-particle states are required so that after dimensional reduction we can sensibly talk about their associated KK modes. These particles need not be perfectly stable or even weakly coupled; a black hole, for our purposes, is a single-particle state. One appealing aspect of this form of the weak gravity conjecture is that it blurs the distinction between black holes and particles. Far out in the charge lattice, the states satisfying the LWGC are extremal black holes. Large black holes are present with all possible charges subject to the extremality bound. At lower masses and charges, the black holes transition to Planck-scale objects and the spectrum of black holes resolves into a discrete spectrum, not well described by semiclassical gravity. At still-lower masses and charges, these objects transition to particles in a low-energy effective field theory. The LWGC is simply an extension of the statement that corrections to the extremality bound reduce the mass of states saturating it.

We note in passing that many or most of the particles satisfying the LWGC will be unstable resonances, except in cases where they are BPS states in a supersymmetric theory, such as the M-theory example discussed above. Nonetheless, these resonances have important consequences upon compactification. For instance, suppose that we have two particles with charges 1 and 2, such that $m_2 = (2+\epsilon) m_1 $ for $\epsilon \ll 1$. In this case, the second particle is unstable. However, upon compactification, the first KK mode of the second particle, with charges $(2,1)$, lies outside the convex hull of the KK modes of the first particle, and is absolutely stable! Thus, despite being a statement largely about unstable resonances, the LWGC has important consequences for the theory.

It is instructive to compare the LWGC to earlier proposals for a ``strong form'' of the WGC. The original paper proposing the WGC~\cite{ArkaniHamed:2006dz} also considered two stronger variants, either 
\begin{enumerate}[(i)]
\item
the particle with least charge should satisfy $Z \ge 1$ or
\item
 the lightest charged particle should satisfy $Z \ge 1$.
\end{enumerate}
 Since~\cite{ArkaniHamed:2006dz} provided several apparent counterexamples for the first proposal, the second has often been considered the ``strong form'' of the WGC. However, extending this ``strong form'' to the case with more than one $U(1)$ is not completely straightforward~\cite{Heidenreich:2015wga}. We cannot simply require that the lightest particle charged under $U(1)_i$ has $Z_i \ge 1$, because this is a basis-dependent statement that requires massless charged particles if enforced in an arbitrary basis. Nor can we merely require that the lightest particle carrying charge under any $U(1)$ has $|\vec{Z}|^2 \ge 1$, as this is a weaker requirement than the convex hull condition.

In principle, sensible generalizations of (ii) to multiple $U(1)$s are possible. For instance, we could require that for $k$ $U(1)$s the lightest $k$ particles which span the charge space should satisfy the convex hull condition \cite{Heidenreich:2015wga}. This would have important consequences for axion inflation. However, the example of M-theory compactified on a torus discussed above does not satisfy this criterion, so this candidate strong form is clearly false. A sounder generalization is the strong form conjectured in \cite{Brown:2015iha} which asks that (ii) is satisfied by the \emph{multiparticle states} in any given direction in the charge lattice.\footnote{However, this criterion does have the unexpected (though not obviously incorrect) feature that adding additional charged particles to an ``allowed'' spectrum can sometimes lead to a ``disallowed'' spectrum.}

Let us evaluate the LWGC in this context. It is easy to see that for a single $U(1)$, the LWGC implies both (i) and (ii), while imposing stricter requirements on the spectrum than either one. To avoid an immediate contradiction, we need to address the claimed counter-examples to (i), the simplest of which is discussed in \S\ref{sec:SO32spinor}. After a careful treatment, accounting for the entire rank-16 Cartan of the $SO(32)$ heterotic gauge group, we find that the heterotic string satisfies the LWGC after all! To understand the tension with~\cite{ArkaniHamed:2006dz}, we note that hypothesis (i) is just as badly behaved as (ii) for more than one $U(1)$. ``The particle with the least charge'' is a basis-dependent question, and we can always choose a pathological basis where a given massive particle with $|Z|^2 \ge 1$ has a small non-zero charge under $U(1)_i$ and a large charge under $U(1)_j$ (or a small charge under many other $U(1)$s), such that $|Z_i| < 1$ and (i) is violated. The heterotic string is indeed a counterexample to (i) in the standard basis for the Cartan of $SO(32)$, but it satisfies the LWGC.

Thus, the LWGC generalizes and extends the ``strong forms'' of the WGC, avoiding the inconsistencies and counter-examples discussed above. The LWGC does not, of course, imply some of the stronger variants of the WGC for multiple $U(1)$s discussed above, but this is fortunate because these variants are ruled out! It does imply the strong form conjectured in \cite{Brown:2015iha}, but strengthens it so that it will remain true after dimensional reduction. We propose the LWGC as the most natural strengthening of the weak gravity conjecture.

\subsection{The LWGC in heterotic string theory}
\label{sec:SO32spinor}

The $SO(32)$ heterotic string has a spinor state at its first massive level \cite{Gross:1985fr,Gross:1985rr,Polchinski:1998rr}, $m^2 = 4/\alpha'$. This spinor carries charge $1/2$ under each $U(1) \cong SO(2)$ subgroup of the $SO(2)^{16}$ Cartan of $SO(32)$. In \cite{ArkaniHamed:2006dz}, this spinor was claimed to be a counterexample to conjecture (i) above. In this section, we show that this is consistent with the LWGC, and indeed that the spectrum of the perturbative heterotic string satisfies the LWGC.

Following the conventions of \cite{Polchinski:1998rr}, the spacetime effective action for the $SO(32)$ heterotic string is 
\be
\frac{1}{2\kappa_{10}^2} \int d^{10} x\sqrt{-g} e^{-2 \Phi} \left(R + 4 \partial_\mu \Phi \partial^\mu \Phi - \frac{\kappa_{10}^2}{g_{10}^2} {\rm Tr}_V\left(|F_2|^2\right)\right),
\ee
where ${\rm Tr}_V$ is the trace in the fundamental representation. We have ${\rm Tr}_V(T^a T^b) = 2 \delta^{ab}$ for the basis of generators $T^a$ including the $SO(2)^{16}$ Cartan discussed above.
If we go to Einstein frame by rescaling by the appropriate power of $e^\phi \equiv e^{\Phi - \Phi_0}$, we obtain
\be
\frac{1}{2 g_s^2 \kappa_{10}^2} \int d^{10}x \sqrt{-g} \left(R - \frac{1}{2} \partial_\mu \Phi \partial^\mu \Phi\right) - \frac{1}{2 g_s^2 g_{10}^2} \int d^{10}x \sqrt{-g} e^{-\phi/2} {\rm Tr}_V\left(|F_2|^2\right),
\ee
so in our conventions we can define 
\begin{align}
8 \pi G_N &= g_s^2 \kappa_{10}^2 = \frac{1}{2} g_s^2 (2\pi)^7 \alpha'^4, &
e^2 &= \frac{1}{2} g_s^2 g_{10}^2 = g_s^2(2\pi)^7 \alpha'^3,
\end{align}
where $e^2$ is the coupling constant associated with any single $U(1) \cong SO(2)$ in the maximal torus. Notice that our dilaton coupling parameter is $\alpha = 1/2$, leading to $\gamma = 1$. So if we restricted our attention to a single $U(1)$, the WGC bound would be
\be
m^2 \leq e^2 q^2/\kappa^2 = \frac{2 q^2}{\alpha'}.
\ee
This is clearly not satisfied for the spinor, with $q=\pm 1/2$ and $m^2 = 4/\alpha'$, hence (i) does not hold for the heterotic string.

In order to compare the nonabelian $SO(32)$ to the WGC, which we have formulated only for abelian gauge groups, we should in principle compactify on a circle and turn on a nontrivial Wilson lines for the Cartan, breaking $SO(32)$ to $U(1)^{16}$ for generic Wilson lines. The black hole region can then be computed in this background and compared with the charge-to-mass ratio of the $SO(32)$ spinor. However, most of these steps can be omitted in practice because, based on the results of~\S\ref{sec:wgc}, the black-hole region of the Cartan will not change upon compactification.\footnote{While the masses of the components of the $SO(32)$ spinor will depend on the Wilson lines---as will the black-hole region---we are free to break $SO(32)$ far below the string scale, where the effect is parametrically subleading.}

Thus, it is sufficient to compute the charge-to-mass vectors of the components of the $SO(32)$ spinor for the Cartan in ten dimensions. 
The spinor at the first excited level  $m^2 = 4/\alpha'$ has $2^{15}$ states with charge vectors (weights)
\be
{\vec q} = \left(\pm \frac{1}{2}, \pm \frac{1}{2}, \ldots \pm \frac{1}{2}\right),
\ee
with the restriction that the number of minus signs is even. Thus, the length of any charge-to-mass vector is
\be
\left|{\vec Z}\right|^2 = \frac{2}{\alpha'} \left|\frac{{\vec q}}{m}\right|^2 = \frac{1}{2} \left|{\vec q}\right|^2 = \frac{1}{2} \times 16 \times \left(\frac{1}{2}\right)^2 = 2.
\ee
We see that the charge-to-mass vectors $\vec{Z}$ for each component of the spinor lie outside the unit ball, hence these points on the charge lattice satisfy the requirements of the LWGC.

It is straightforward to extend this argument to a proof of the LWGC for the perturbative heterotic string. The charge lattice of the $SO(32)$ heterotic string consists of all charge vectors of the form:
\begin{align}
{\vec q} &= \left(q_1, q_2, \ldots q_{16}\right), & \mathrm{or} & & {\vec q} & = \left(q_1 + \frac{1}{2}, \ldots, q_{16} + \frac{1}{2}\right)
& \mathrm{with} & & q_i &\in \mathbb{Z}, & \sum_i q_i &\in 2 \mathbb{Z}. 
\end{align}
This lattice is even, in that $|q|^2 \in 2 \mathbb{Z}$ for any $\vec{q}$ in the lattice.
In the bosonic construction of the heterotic string, the mass-shell condition is
\be
\frac{\alpha'}{4} m^2 = N_L + \frac{1}{2} \left|{\vec q}\right|^2 - 1 = N_R - \frac{1}{2},
\ee
where $N_{L,R}$ are the occupation number of the left and right-moving oscillators, with $N_L$ a non-negative integer and $N_R$ a positive half-integer. Since for any choice of $N_L \ge 0$ and $\vec{q} \ne 0$, $N_R$ can be chosen to satisfy the level-matching condition, the lightest state with a given $\vec{q} \ne 0$ has
\be
m^2 = \frac{2}{\alpha'} \left(\left|{\vec q}\right|^2 - 2\right).
\ee
 Thus, there is always a state in the spectrum of the perturbative heterotic string for any allowed charge ${\vec q}$ that has
\be
\left|{\vec Z}\right|^2 = \frac{2}{\alpha'} \left|\frac{{\vec q}}{m}\right|^2 = \frac{\left|{\vec q}\right|^2}{\left|{\vec q}\right|^2 -2 } > 1.
\ee
This shows that the LWGC is true for the perturbative heterotic $SO(32)$ string, with the particles of lowest mass for a given charge becoming increasingly superextremal for smaller charges. In fact, the same argument goes through for the perturbative heterotic $E_8 \times E_8$ string, mutatis mutandis.\footnote{Similar arguments were given in~\cite{ArkaniHamed:2006dz} for a $T^6$ compactification of the heterotic string, but in support of the ordinary WGC rather than the LWGC.}

A second counterexample 
to the conjecture (i) of~\cite{ArkaniHamed:2006dz}
is based on fractionally charged strings in certain string compactifications \cite{Wen:1985qj}. A similar argument that (i) is violated but the LWGC is satisfied may be possible for these states as well.
We expect that this example will also provide an interesting case study for the Single-EFT Consistency Criterion that we proposed in \cite{Heidenreich:2015wga}, since from the low-energy effective theory point of view it involves nonminimal Dirac quantization of electric and magnetic charges. We leave a close consideration of this example for future work, since a detailed assessment of the bound seems to depend on details of the compactification. 

\section{The Weak Gravity Conjecture for axions}
\label{sec:axions}

In this section we will explore the degree to which the Weak Gravity Conjecture can be extended to axion fields. The original paper on the Weak Gravity Conjecture \cite{ArkaniHamed:2006dz} suggested an analogy in which the axion is a zero-form gauge field with coupling $1/f$ (where $f$ is the axion decay constant), objects charged under the axion are instantons, and the ``tension'' of these instantons is their action $S_{\rm inst}$. At a superficial level, this analogy is appealing, especially in light of its possible application to theories of axion inflation. However, as we noted in the discussion below (\ref{eq:generalWGC}), the situation is not so simple: a na\"ive extrapolation of the conjectured bound fails in the case of zero dilaton coupling. There is still hope, however, because the bound could be nontrivial in the presence of a dilaton coupling. Indeed, axions in string theory generally arise (at least in some duality frame) from integrating a $p$-form gauge field ($p > 0$) over a cycle, and so the axion will come with a coupling to the dilaton-like field that controls the volume of the cycle. We will pursue this idea from two points of view. First, in a theory of axions coupled to dilatons, we will construct gravitational instanton solutions and show that they obey an extremality condition similar to that obeyed by black holes. For a certain range of dilaton couplings $\alpha$ this extremality condition is precisely what we would find by na\"ively extrapolating (\ref{eq:generalWGC}) to the case $p = 0$. Second, we will consider axions that arise from compactification of theories with higher rank $p$-forms, and show that the instantons arising from wrapped objects match in a well-defined way onto the gravitational instantons. This lends support to the idea that the Weak Gravity Conjecture applies to axions. It also suggests that the gravitational instanton solutions correspond to an approximate effective description of effects arising in the underlying higher-dimensional theory from wrapped Euclidean worldvolumes of charged objects, rather than a completely independent effect. 

Before explaining our results, we will first briefly review the way that axion potentials arise from wrapped objects.

\subsection{Reminder: axions, loops, and instantons}
\label{subsec:review}

Axions arise in contexts like extranatural inflation \cite{ArkaniHamed:2003wu} and string theory from integrating gauge fields over cycles. The axion obtains a potential as a semiclassical instanton effect arising from a Euclidean worldvolume wrapped around the compact cycle. Because we will be studying gravitational instanton solutions in relation to such wrapped Euclidean objects, it is useful to  review this picture of the axion potential.

For simplicity and concreteness we will discuss the simplest case, a four-dimensional axion $A_0(x) = \oint_0^R dx^5 A_5(x,x^5)$ arising from a Wilson loop around a compact fifth dimension, which inherits a discrete $A_0 \to A_0 + 2\pi$ shift symmetry from large gauge transformations on the circle. A particle of fermion number $F$, mass $m$, and charge $q$ in 5d gives rise to an effective potential for the axion, which is a sum over $n$-instanton terms:
\begin{equation}
\delta V(A_0) = \frac{3(-1)^F}{4 \pi^2} \frac{1}{(2 \pi R)^4} \displaystyle\sum_{n\in \mathbb{Z}} c_n e^{-2 \pi R m n} \text{Re}(e^{i q n A_0}),
\label{extranaturaleq}
\end{equation}
where \cite{ArkaniHamed:2003wu,Hosotani:1983xw,Cheng:2002iz,delaFuente:2014aca}
\begin{equation}
c_n(2 \pi R m) = \frac{(2 \pi R m)^2}{3 n^3} + \frac{2 \pi R m}{n^4} + \frac{1}{n^5}.
\label{prefactoreq}
\end{equation}
This potential can be thought of as the Casimir energy associated with the compact cycle, and is often discussed as a one-loop effect due to the axion coupling to a tower of Kaluza-Klein modes. In this case, the Poisson summation formula can be used to reinterpret the loop computation as a sum over winding numbers \cite{Cheng:2002iz}.

For our purposes, a more useful approach follows the Casimir effect computation in Appendix A of \cite{ArkaniHamed:2007gg}. The computation, for a scalar field of mass $m$ in $d$ dimensions compactified on a circle of radius $R$ down to $d-1$ dimensions, writes the expectation value of the stress-energy tensor in terms of a differential operator acting on the propagator. Because of the periodic identification of the compactified direction, the Green's function involves a sum over images of the particle separated by distances $2\pi R n$ in the compact direction:
\be
V(r) = 2 \sum_{n \neq 0} e^{i n A_0} \left.\frac{\partial G(r^2)}{\partial r^2}\right|_{r = 2 \pi R n}.
\ee 
Each term in this sum involves propagation $n$ times around the circle and can be thought of as a Euclidean worldline instanton wrapping the circle $n$ times. The Green's function is a standard free-particle propagator. It is straightforward to derive (\ref{extranaturaleq}) and (\ref{prefactoreq}) from the usual textbook expressions for the propagator. To make the connection to instanton effects more explicit, it is useful to rewrite the propagator in terms of Schwinger proper time:
\be
G(x) = \int \frac{d^d k}{\left(2\pi\right)^d} \frac{e^{i k \cdot x}}{k^2 + m^2} = \int \frac{d^d k}{\left(2\pi\right)^d}  \int_0^\infty d\tau e^{i k \cdot x} e^{-\tau(k^2 + m^2)}.
\ee
We can now integrate over $k$. This leaves an integral over $\tau$ which is straightforward when $m = 0$ and is well-approximated by a steepest-descent estimate when $m > 0$, which simplifies greatly for $|m x| \gg d$:
\be
G(x^2) = \int_0^\infty d\tau \frac{1}{\left(4 \pi \tau\right)^{d/2}} e^{-\frac{x^2}{4\tau}} e^{-\tau m^2} \approx \begin{cases} \left(\frac{m^2}{4 \pi^2 x^2}\right)^{(d-1)/4}\frac{e^{-m x}}{2m},& \text{if } |m x| \gg d\\
 \frac{\Gamma(d/2-1)}{4 \pi^{d/2}} \left(\frac{1}{x^2}\right)^{d/2 - 1},& \text{if } m = 0.
 \end{cases}
\ee
Notice the exponential factor $e^{-mx} = e^{-2\pi n m R}$, as expected for an instanton arising from a wrapped worldline. This computation sheds light on the prefactor $1/n^5$ in (\ref{prefactoreq}), which simply comes from the power-law dependence of the five-dimensional propagator on the distance propagated. This factor played an important role in \cite{delaFuente:2014aca} by suppressing higher harmonics and leading to the ``small-action loophole.''

The chief reason for discussing the calculation in this way is to bring out the similarity to the way one would calculate the action of a wrapped Euclidean string or D-brane. Here $\tau$ is the proper time on the worldline. A calculation for a string would generalize to a double integral over both worldsheet coordinates $\tau$ and $\sigma$. The prefactor $(4 \pi \tau)^{-d/2}$ is the functional determinant over the fields describing the embedding of the worldline into spacetime. This, too, will generalize to include the dynamical fields on a wrapped string or brane. For objects of finite tension, the steepest-descent estimate will produce the expected factor $\exp(-n \oint_\Sigma T)$. The prefactor of an $n$-instanton effect is again expected to behave as a power law at large $n$, due to the cost of propagating even a light string or brane over an extended distance. It would be interesting to fill in more details of the estimated prefactors for higher-dimensional objects.

\subsection{Gravitational instantons}

In addition to black holes and black branes, there are Euclidean solutions to the Einstein-Maxwell-dilaton (or Einstein-axion-dilaton) theory, known as ``gravitational instantons.'' These solutions differ from black hole and black brane solutions for two related reasons:
\begin{enumerate}
\item
They are solutions to the Euclidean (rather than the Lorentzian) action.
\item
Due to the absence of a time coordinate, there is no horizon.
\end{enumerate}
Instead, gravitational instanton solutions come in three classes:
\begin{enumerate}
\item
Solutions with a singular core.
\item
Solutions with a flat metric (the ``extremal'' case).
\item
Wormhole solutions, with two different asymptotic regions connected by a smooth throat.
\end{enumerate}
Of these possibilities, the wormhole case has received the most attention~\cite{Giddings:1987cg,Lee:1988ge,Giddings:1989bq,Coleman:1989zu,Abbott:1989jw,Kallosh:1995hi,ArkaniHamed:2007js}, as it is the only example that is smooth and non-trivial. However, the interpretation of this solution as an instanton is problematic.\footnote{See e.g.~\cite{ArkaniHamed:2007js}. In the context of axion inflation, wormhole solutions have previously been studied by~\cite{Montero:2015ofa,Bachlechner:2015qja} and criticized by~\cite{Brown:2015iha,Brown:2015lia}.}

In particular, our focus in the present paper is on instantons which contribute to the axion potential. To do so, the instanton must carry charge under the Hodge-dual $D-2$ form. This is true of each end of the Euclidean wormhole considered separately, but taken together the net charge of the two ends is zero. Thus, the Euclidean wormhole is not a charged object \emph{per se}, but rather a conduit by which charge could flow from one place to another. Since the net charge is unchanged in this process, the wormhole cannot contribute to the axion potential, regardless of its role in quantum gravity.

Conversely, the cored solution carries a net charge at the expense of a curvature singularity at its center. Since there can be no horizon, this singularity is naked.  Indeed, the appearance of a singularity---at least in the flux density---is ensured by charge conservation and spherical symmetry. However, as we will show, the instanton action is finite and computable despite this, as the contributions near the singularity are negligible.

Just as a many-particle state can collapse to form a black hole, we propose that an instanton with a large charge---for which gravitational backreaction cannot be neglected---will collapse into a cored gravitational instanton of the type we consider here. (Collapsing into a Euclidean wormhole is not possible, due to the differing topologies.) Thus, this type of gravitational instanton (unlike the Euclidean wormhole) is analogous to a black hole, and should play a similar role in the weak gravity conjecture.

We begin by reviewing the different types of gravitational instanton solutions, followed by a discussion of the instanton action and how it changes upon dimensional reduction. Several of our results have appeared in some form in the literature \cite{Gutperle:2002km,Bergshoeff:2004fq,Bergshoeff:2004pg}. We find it useful to rederive the solutions with our preferred conventions while emphasizing the aspects we find physically important.

\subsection{Instanton solutions}

Gravitational instantons are rotationally invariant solutions of the Euclidean action:
\be
S_E = \frac{1}{2\kappa_d^2} \int d^d x \sqrt{g} \left(-{\cal R} + \frac{1}{2} (\nabla \phi)^2 \right) + \frac{1}{2e_{d-2;d}^2}  \int d^d x \sqrt{g} e^{-\alpha \phi} F_{d-1}^2 \,.\label{eq:Euclideanaction}
\ee
where we have chosen to work with a $d-2$ form $A_{d-2}$---under which the solutions are magnetically charged---instead of the Hodge dual axion $A_0$.
The Einstein equations together with rotational invariance imply that the angular components of the Ricci tensor vanish, hence the metric takes the form
\begin{equation}
d s^2 = \left(1+\frac{C}{r^{2(d-2)}}\right)^{-1} dr^2+r^2 d \Omega_{d-1}^2 \,.
\end{equation}
for some $C$. For $C<0$, there is a coordinate singularity at $r = r_w \equiv (-C)^{\frac{1}{2(d-2)}}$. Defining
\be
\frac{r^{d-2}}{r_w^{d-2}} \equiv \cosh[(d-2)u]\,,
\ee
we obtain
\be
ds^2 = r^2 (du^2+d\Omega_{d-1}^2)\,,
\ee
which is smooth, where $r>r_w$ corresponds to $u>0$. There is a reflection symmetry $u \to -u$, so the geometry is that of a two-sided wormhole, with topology $\mathbb{R}\times S^{d-1}$ and minimum radius $r_w$. Conversely, for $C>0$ the geometry is smooth for $r>0$, but there is a curvature singularity at $r=0$, where $\mathcal{R} \sim r^{-2(d-1)}$. For $r\ll r_0 \equiv C^{\frac{1}{2(d-2)}}$, the volume of the angular $S^{d-1}$ shrinks rapidly, and the space closes off, rather than opening up as it does in the Euclidean wormhole. For $C=0$, the geometry is flat.

For $C>0$, the dilaton profile is
\be \label{eqn:dilatonprofile}
e^{\alpha \phi} = \frac{1}{\sinh^2 \psi_0} \sinh^2 \left[\psi_0 + \frac{\alpha}{\alpha_0} \sinh^{-1} \frac{r_0^{d-2}}{r^{d-2}} \right]\,,
\ee
where we have fixed $\phi \to 0$ as $r\to \infty$, and $\alpha_0$ is the critical coupling
\be
\alpha_0 \equiv \sqrt{\frac{2 (d-2)}{d-1}} \,.
\ee
Not coincidentally, this is the same as the radion coupling $\beta_{d-2;d}$ from~(\ref{eqn:radioncoupling}), which will play an important role in our analysis.

The integration constant $\psi_0$ in~(\ref{eqn:dilatonprofile}) depends on $C$ and the charge of the solution. In particular
\be
C = r_0^{2(d-2)} = \frac{\kappa^2 4 \pi^2 {\tilde Q}^2 \sinh^2 \psi_0}{e^2 V_{d-1}^2 (d-1)(d-2)}\,,
\ee
with $V_{d-1}$ and ${\tilde Q}$ defined as in equations (\ref{eq:sphereV}) and (\ref{eq:Qmag}).
We restrict to $\psi_0 \ge 0$ to avoid a singularity in the dilaton profile at finite radius. 
The asymptotic behavior at large and small $r$ is then
\be
\phi =\begin{cases} \frac{\sqrt{2} \kappa 2\pi |{\tilde Q}| \cosh \psi_0}{e V_{d-1} (d-2) r^{d-1}}+\ldots & r \gg r_0 \\[1ex]
       \frac{2}{\alpha_0} \log \frac{2 r_0^{d-2}}{r^{d-2}}+\frac{2}{\alpha} \left[\psi_0 - \log (2 \sinh \psi_0)\right]+\ldots & r \ll r_0
       \end{cases}
\ee
Notice that the dilaton cannot be truncated. The only non-trivial solutions with constant $\phi$ are wormholes.

\subsection{The instanton action}

We now evaluate the action of the cored gravitational instantons described in the previous section. To do so consistently, we need to include the appropriate boundary terms, in particular the Gibbons-Hawking-York surface term:
\begin{multline}
S_E = \frac{1}{2\kappa_d^2} \int_{\mathcal{M}} d^d x \sqrt{g} \left(-{\cal R} + \frac{1}{2} (\nabla \phi)^2 \right) + \frac{1}{2e_{d-2;d}^2}  \int_{\mathcal{M}} d^d x \sqrt{g} e^{-\alpha \varphi} F_{d-1}^2\\
-\frac{1}{\kappa_d^2} \oint_{\partial\mathcal{M}} d^{d-1} x \sqrt{g} (K_{\mathcal{M}}-K_{\mathcal{M}}^{(0)}) \,, \label{eq:Euclideanaction2}
\end{multline}
where $K_{\mathcal{M}}$ is the extrinsic curvature of $\partial \mathcal{M}$ within $\mathcal{M}$ and $K_{\mathcal{M}}^{(0)}$ is the extrinsic curvature of $\partial \mathcal{M}$ embedded in flat space such that the pullback metric is the same. Note that there is no boundary term associated to $F_{d-1}$. The Hodge-dual action for the corresponding axion does have an additional boundary term, but this term cancels the boundary term induced by Hodge-duality~\cite{Bergshoeff:2004fq}.

The Euclidean action~(\ref{eq:Euclideanaction2}) can be simplified by imposing the equations of motion. We obtain:
\be
S_E = -\frac{1}{\kappa^2} \oint_{\partial\mathcal{M}} d^{d-1} x \sqrt{g} \left(\frac{1}{\alpha} n^{\mu}\nabla_\mu \phi+K_{\mathcal{M}}-K_{\mathcal{M}}^{(0)}\right) \,, \label{eq:EucActionOnShell}
\ee
where $n^{\mu}$ is the outward directed unit normal and $K_{\mathcal{M}} = \nabla_{\mu} n^{\mu}$. Thus, the on-shell action reduces to boundary terms. Consider a surface of fixed radius $r$. The outward pointing unit normal is
\be
n = \left[1+\frac{r_0^{2(d-2)}}{r^{2(d-2)}}\right]^{1/2} \frac{\partial}{\partial r}\,.
\ee
Using this, we obtain
\be
n^{\mu} \nabla_{\mu} \phi = -\frac{\kappa 2\sqrt{2}\pi |{\tilde Q}| \sqrt{e^{-\alpha \phi} + \sinh^2 \psi_0}}{e V_{d-1}  r^{d-1}} \;\;,\;\; K_{\mathcal{M}} - K_{\mathcal{M}}^{(0)} = \frac{d-1}{r} \left[\left(1+\frac{r_0^{2(d-2)}}{r^{2(d-2)}}\right)^{1/2}-1\right]\,,
\ee
which gives the boundary term
\be
S(r)=\frac{2\sqrt{2} \pi |{\tilde Q}| \sqrt{e^{-\alpha \phi} + \sinh^2 \psi_0}}{\alpha e \kappa}-\frac{d-1}{\kappa^2} \left[\left(r^{2(d-2)}+r_0^{2(d-2)}\right)^{1/2}-r^{d-2}\right] V_{d-1} \,.
\ee
The integrated on-shell action is then
\be
S_E = S(\infty)-S(0) = \frac{2\sqrt{2} \pi |{\tilde Q}|}{e \kappa} \left[\frac{1}{\alpha} e^{-\psi_0} + \frac{1}{\alpha_0} \sinh \psi_0 \right]\,.
\ee

Finally, we minimize the action as a function of $\psi_0 \ge 0$ to find the dominant instanton for any fixed $|{\tilde Q}|$. For $\alpha \ge \alpha_0$, the minimum occurs at $\psi_0 =0$, whereas for $\alpha < \alpha_0$, the minimum is at
\be
\psi_0 = \frac{1}{2} \log \left(\frac{2 \alpha_0}{\alpha} - 1\right) \,.
\ee
Thus, the minimum instanton action is
\be \label{eqn:Sextremal}
S_{\rm min} = \frac{2\sqrt{2} \pi |{\tilde Q}|}{e \kappa} \times \begin{cases} \frac{1}{\alpha} & \alpha \ge \alpha_0 \\ \frac{1}{\alpha_0} \sqrt{\frac{2 \alpha_0}{\alpha} - 1} & \alpha < \alpha_0 \end{cases} \,.
\ee
Note that $S_{\rm min}/|Q|$ is a monotonically decreasing function of $\alpha$. If we interpret~(\ref{eqn:Sextremal}) as an extremality bound for these ``black instantons'' (c.f.~\cite{Bergshoeff:2004fq,Bergshoeff:2004pg})---where the instanton action now plays the role of the black hole mass---then the corresponding weak gravity conjecture for instantons obeys the same kind of monotonicity that we saw for extended objects: stabilizing the dilaton always weakens the bound.

We derived the magnetic extremality bound (\ref{eq:magneticextremality}) in the case $1 \leq p \leq d-3$, but if we na\"ively extrapolate it to the case $p = d-2$, identifying the tension $T$ with the instanton action $S_{\rm inst}$, we obtain
\be
\gamma g_{d-2;d}^2 {\tilde Q}^2 = \frac{2}{\alpha^2} \frac{4\pi^2}{e_{d-2;d}^2} {\tilde Q}^2 \leq \kappa^2 S_{\rm inst}^2.
\ee
In other words, $S_{\rm inst} \geq \frac{2\pi \sqrt{2}  {\tilde Q}}{\alpha e \kappa}$, precisely agreeing with (\ref{eqn:Sextremal}) in the case $\alpha \geq \alpha_0$. This suggests that gravitational instanton solutions play a role closely analogous to black holes, providing support for the notion that the Weak Gravity Conjecture can be extrapolated to the cases $p = 0$ and $p = d-2$ in a well-defined way (at least for a certain range of dilaton couplings $\alpha$). We will find further evidence for this claim by considering axions arising from dimensional reduction.

\subsection{Dimensional reduction}

In the above discussion, we have studied the action of instantons within a $d$-dimensional effective theory. However, we know many examples in which instanton effects are best understood as wrapped Euclidean worldvolumes of charged objects in higher dimensional theories, as we reviewed in \S \ref{subsec:review}. The recent interest in gravitational instantons in the context of the Weak Gravity Conjecture \cite{Montero:2015ofa,Bachlechner:2015qja,Brown:2015lia} motivates the question: when are the instanton effects we have discussed above the {\em same} as those arising from wrapped worldvolumes in higher dimensions?

To approach this problem, consider the case of a $D=d+1$ dimensional theory with a $d-2=D-3$ form gauge field. This theory has charged black hole solutions and can be compactified on a circle of radius $R$ to $d$ dimensions, yielding a theory with an axion-like field that has Euclidean instanton solutions as described above. Our question is, do these instanton solutions lift to higher-dimensional Euclidean black hole solutions in such a way that the Euclidean instanton action is $S_E = 2\pi R M_{\rm ADM}$, where $M_{\rm ADM}$ is the ADM mass of the black hole?

In fact, this question can be answered in complete generality, without reference to a particular black hole solution. Consider a black hole spacetime in $D=d + 1$ dimensions with an ADM
decomposition:
\begin{equation}
  d s^2 = - N^2 d t^2 + h_{a b}  (d y^a +\mathcal{N}^a d
  t)  (d y^b +\mathcal{N}^b d t)
\end{equation}
Upon reducing along the time direction, $N$ becomes the radion and
$\mathcal{N}^a$ becomes the graviphoton, whereas the dimensionally reduced
Einstein-frame metric is
$  \tilde{h}_{a b} = N^{\frac{2}{d - 2}} h_{a b}\,.$
The ADM mass of the solution is:
\begin{equation} \label{eqn:ADMmass}
  M_{\rm ADM} = - \frac{1}{\kappa_{D}^2} \oint_{\partial \Sigma_t}
  d^{D - 2} y \sqrt{h} N (K_{\Sigma} - K^{(0)}_{\Sigma})
\end{equation}
where $\Sigma_t$ is a surface of constant time.
We  rewrite this in terms
of the $d$-dimensional metric $\tilde{h}_{a b}$. 
 Under a conformal
transformation
  $\tilde{g}_{\mu \nu} = e^{2 \omega} g_{\mu \nu}$, 
we obtain
 $ \tilde{n}^{\mu} = e^{- \omega} n^{\mu}$.
Therefore, the extrinsic curvature in $d$ dimensions transforms into:
\begin{equation}
  K = \nabla_{\mu} n^{{\mu}} =  \frac{e^{\omega
  d}}{\sqrt{\tilde{g}}} \partial_{{\mu}} \left( e^{- \omega (d - 1)} 
  \sqrt{\tilde{g}}  \tilde{n}^{{\mu}} \right) = e^{\omega}  (\tilde{K} - (d - 1)  \tilde{n}^{{\mu}} 
  \tilde{\nabla}_{{\mu}} \omega) \,,
\end{equation}
whereas the reference curvature is simply rescaled
  $K^{(0)} = e^{\omega}  \tilde{K}^{(0)}$.
Thus,
\begin{equation} \label{eqn:ADMred}
  M_{\rm ADM} = - \frac{1}{\kappa_D^2}  \oint_{\partial \Sigma} d^{d -
  1} y \sqrt{\tilde{h}}  (\tilde{K}_{\Sigma} - \tilde{K}^{(0)}_{\Sigma}) +
  \frac{1}{\kappa_D^2}  \oint_{\partial \Sigma} d^{d - 1} x
  \sqrt{\tilde{h}}  \left( \tilde{n}^{{\mu}}  \tilde{\nabla}_{{\mu}}
  \log N^{\frac{d - 1}{d - 2}} \right)
\end{equation}
where $\Sigma$ is the $d$-dimensional Euclidean space, and the first term is
just the usual Gibbons-Hawking-York surface term associated to the Euclidean
Einstein-Hilbert term.

We compare this with the on-shell Euclidean action~(\ref{eq:EucActionOnShell}):
\be \label{eq:EucActionOnShell2}
S_E = -\frac{1}{\kappa_d^2} \oint_{\partial \Sigma} d^{d-1} x \sqrt{g} \left(\frac{1}{\alpha_{d-2;d}} n^{\mu}\nabla_\mu \rho+K_{\Sigma}-K_{\Sigma}^{(0)}\right) \,.
\ee
where $\rho$ is the effective dilaton in $d$ dimensions. To relate this to~(\ref{eqn:ADMred}), we use the results of~\S\ref{subsec:dred-pconst} with $\log N^2 = -\lambda$, hence
\be
\log N^{\frac{d-1}{d-2}} = -\frac{1}{\alpha_0^2} \lambda = -\frac{1}{\alpha_0} \hat{\lambda} = -\frac{\rho}{\alpha_{d-2;d}} + \frac{\alpha_{d-2;D}}{\alpha_0 \alpha_{d-2;d}} \sigma
\ee
where $\alpha_0 = \beta_{d-2;d}$ is the radion coupling and we have expressed $\hat \lambda$ in terms of the conventionally normalized field $\rho$ that couples to the field strength and $\sigma$ that does not. The equation of motion $\nabla^2 \sigma = 0$ ensures that the flux integral of $\nabla\sigma$ over the boundaries vanishes. Thus, since $\kappa_d^2 = \kappa_{D}^2 / (2 \pi R)$,
\begin{equation} \label{eqn:mass-action-reln}
  S_E = (2 \pi R) M_{\rm ADM}\,,
\end{equation}
independent of the details of the black hole solution.

In fact, the apparent generality of~(\ref{eqn:mass-action-reln}) is somewhat misleading. The ADM mass, (\ref{eqn:ADMmass}), is evaluated in the $D$-dimensional Lorentzian black hole spacetime, and only receives contributions at spatial infinity. Conversely, the instanton action~(\ref{eq:EucActionOnShell2}) is evaluated on the $d$-dimensional instanton solution in Euclidean signature---which has no horizon---and can receive contributions at other boundaries. When other boundaries contribute, (\ref{eqn:mass-action-reln}) will not hold. For instance, there is always another boundary contribution for Euclidean wormholes---the far end of the wormhole\footnote{In fact, for $\alpha_{d-2;d} \ge \alpha_0$ (as for an unstabilized radion) the dilaton profile within the wormhole blows up at finite distance from the center, further complicating the interpretation of this solution.}---and this formula does not apply. Conversely, in the ``black instanton'' case considered above the inner boundary term is proportional to $\sinh \psi_0$, whereas $\alpha_{d-2;d} \ge \alpha_0$ by~(\ref{eqn:psame_alpha}), so that the minimum action instanton has $\psi_0 = 0$, and~(\ref{eqn:mass-action-reln}) holds.

\section{Conclusions}
\label{sec:outlook}

We have seen that the convex hull condition (CHC) implied by the WGC is more subtle than has been appreciated before. The precise bound depends on the moduli fields of the theory. In all examples we have studied, integrating out dilaton fields weakens the CHC bound, indicating that the WGC bound grows weaker as one flows to the IR. However, the appearance of new, Kaluza-Klein $U(1)$ gauge groups under compactification implies that satisfying the CHC in a given theory is not a guarantee that the condition will still be satisfied after compactifying. This imposes further constraints on the higher dimensional theory, enforcing either a larger minimal charge-to-mass ratio than one would have na\"ively expected from the WGC or else necessitating the existence of additional charged particles.  

We have presented two independent lines of evidence indicating the WGC can indeed be extended to axions as hypothesized in \cite{ArkaniHamed:2006dz} once a dilaton coupling to the axion is turned on.  Additionally, our work suggests that extremal gravitational instantons cannot be used to satisfy the axionic WGC in the same way that extremal black holes cannot be used to satisfy the WGC for 1-form gauge fields. In the case of the Lattice WGC, extremal black holes can play a role for large charges in the charge lattice but points of small charge require a lighter particle or string state. The analogous conjecture for axions is that small-charge instantons must exist that are not semiclassical gravitational instanton solutions.

We see several directions for future progress. The study of combinations of arbitrary $U(1)$ gauge fields with Kaluza-Klein gauge fields has proved to be interesting. Adding magnetic charges to this picture, e.g.~by constructing solutions that give additional charges (magnetic or electric) to the KK monopole, would be an interesting exercise that might lead to new physical insights. We have also argued in \S \ref{sec:SO32spinor} that perturbative heterotic string theory satisfies a very strong version of the WGC---the Lattice Weak Gravity Conjecture, requiring a superextremal or extremal particle for each allowed point in the charge lattice. We plan to explore a wider range of string theories to understand whether the Lattice WGC continues to be true in settings beyond the heterotic string. If so, this would have important consequences for models of axion inflation.

The most important task is to put the WGC itself on a more rigorous footing. There is a great deal of circumstantial evidence for the conjecture, and in this paper we have seen that appropriate versions of the conjecture can pass a new battery of tests arising from compactification. But, to date, there is no very compelling argument for why the WGC must be true. The statement that it is needed to avoid a plethora of stable black hole states is intriguing, but (unlike for the case of arguments against global symmetries) these stable black holes are spread over a wide range of masses and not in obvious conflict with general principles like entropy bounds (see \cite{Banks:2006mm}, however, for an attempt to construct an argument for a bound parametrically resembling the WGC based on the Covariant Entropy Bound). Thus, there is a strong need for either sharper arguments based on black hole thermodynamics or a new approach to deriving the WGC from general principles. Along these lines, attempts to derive bounds on the low-energy effective action of quantum gravity from analyticity and unitarity of scattering amplitudes are noteworthy \cite{Cheung:2014ega,Bellazzini:2015cra}, but so far only partially successful: what they constrain are combinations of the mass and charge of particles in the theory together with unknown ultraviolet-sensitive coefficients of higher-dimension operators. Without either a refined argument or some control over these ultraviolet contributions, it is unclear if such arguments can prove the desired result. We could also hope that an appropriate AdS generalization of the WGC could be proven using conformal field theory techniques. If nothing else, CFTs provide a new catalogue of examples to check, an approach that we have pursued in some detail and will report on in a separate publication. (While this work was in progress a related preprint appeared \cite{Nakayama:2015hga}.)

The Weak Gravity Conjecture offers a hope of linking phenomenological questions, like the presence or absence of tensor modes in the CMB, with deep general questions regarding the nature of quantum gravity. The body of evidence in favor of the conjecture and its consistency is steadily growing, but we have also shown that the conjecture has unexpected subtleties. We feel certain that further exploration will be rewarding.

\section*{Acknowledgments}
We thank Thomas Bachlechner, Cody Long, Liam McAllister, and Cumrun Vafa for discussions or correspondence. BH is supported by the Fundamental Laws Initiative of the Harvard Center for the Fundamental Laws of Nature. The work of MR is supported in part by the NSF Grant PHY-1415548. TR is supported in part by the National Science Foundation under Grant No.~DGE-1144152. MR's
work was supported in part by the National Science Foundation under Grant No.~PHYS-1066293 and the hospitality of the Aspen Center for Physics.

\bibliography{ref}
\bibliographystyle{utphys}
\end{document}